\documentclass[journal]{IEEEtran}
\usepackage{graphicx}
\usepackage{epsfig}
\usepackage{amsmath}
\usepackage{amssymb}
\usepackage{cite}
\usepackage{color}
\usepackage{multirow}
\usepackage{subfigure}

\begin{document}

\title{Deep Neural Network for Fast and Accurate Single Image Super-Resolution via Channel-Attention-based Fusion of Orientation-aware Features}

\author{Du~Chen*,
        Zewei~He*,
        Yanpeng~Cao$\dagger$,~\IEEEmembership{Member,~IEEE},
        Jiangxin~Yang,
        Yanlong~Cao,
        Michael~Ying~Yang,~\IEEEmembership{Senior Member,~IEEE},
        Siliang~Tang,
        and~Yueting~Zhuang
\thanks{* The first two authors contributed equally to this work.}
\thanks{This work was supported in part by National Natural Science Foundation of China (No. 51605428, 51575486).}
\thanks{D.~Chen, Z.~He, Y.~Cao, J.~Yang and Y.~Cao are with State Key Laboratory of Fluid Power and Mechatronic Systems and Key Laboratory of Advanced Manufacturing Technology of Zhejiang Province, School of Mechanical Engineering, Zhejiang University, Hangzhou, 310027, China (e-mail: caoyp@zju.edu.cn).}
\thanks{M.~Y.~Yang is with the Scene Understanding Group in University of Twente. (e-mail: michael.yang@utwente.nl).}
\thanks{S.~Tang and Y.~Zhuang are with College of Computer Science and Technology, Zhejiang University, Hangzhou, 310027, China (e-mail: siliang@zju.edu.cn).}
\thanks{$\dagger$ Corresponding authors: Yanpeng~Cao.}
}

\markboth{Journal of \LaTeX\ Class Files,~Vol.~X, No.~X, November~2019}
{Shell \MakeLowercase{\textit{et al.}}: Bare Demo of IEEEtran.cls for IEEE Journals}

\maketitle

\begin{abstract}
Recently, Convolutional Neural Networks (CNNs) have been successfully adopted to solve the ill-posed single image super-resolution (SISR) problem. A commonly used strategy to boost the performance of CNN-based SISR models is deploying very deep networks, which inevitably incurs many obvious drawbacks (e.g., a large number of network parameters, heavy computational loads, and difficult model training). In this paper, we aim to build more accurate and faster SISR  models via developing better-performing feature extraction and fusion techniques. Firstly, we proposed a novel Orientation-Aware feature extraction and fusion Module (OAM), which contains a mixture of 1D and 2D convolutional kernels (i.e., $5\times1$, $1\times5$, and $3\times3$) for extracting orientation-aware features.  Secondly, we adopt the channel attention mechanism as an effective technique to adaptively fuse features extracted in different directions and in hierarchically stacked convolutional stages. Based on these two important improvements, we present a compact but powerful CNN-based model for high-quality SISR via Channel Attention-based fusion of Orientation-Aware features (SISR-CA-OA).  Extensive experimental results verify the superiority of the proposed SISR-CA-OA model, performing favorably against the state-of-the-art SISR models in terms of both restoration accuracy and computational efficiency. The source codes will be made publicly available.
\end{abstract}


\IEEEpeerreviewmaketitle

\section{Introduction}

\IEEEPARstart{S}{ingle} image super-resolution (SISR) restores a high-resolution (HR) image containing abundant details and textures based on its low-resolution (LR) version. It provides an effective technique to increase the spatial resolution of optical sensors and thus has attracted considerable attention from both the academic and industrial communities. In last decades, many machine learning SISR algorithms have been developed, such as sparse coding \cite{yang2008image,yang2010image}, local linear regression \cite{timofte2014a+} and random forest \cite{schulter2015fast}. However, SISR remains a challenging ill-posed problem because one specific LR input can correspond to many possible HR versions, and the mapping space is too vast to explore.

In recent years, Convolutional Neural Networks (CNNs) have been successfully adopted to solve the SISR problem by implicitly learning the complex nonlinear LR-to-HR mapping relationship based on numerous LR-HR training image pairs. SRCNN~\cite{dong2014learning} proposed a three-layer CNN model to learn the nonlinear LR-to-HR mapping function.  It is the first time that deep learning technique is applied to tackle the SISR problem. Compared with the many traditional machine-learning-based SISR methods, the lightweight SRCNN model achieved significantly improved image restoration results. Since then, many CNN-based models have been proposed to achieve more accurate SISR results \cite{kim2016accurate, tai2017image, tong2017image, tai2017memnet, huang2017densely, hu2019channel, NTIRE2017SRreport, NTIRE2018SRreport, NTIRE2019SRreport}. Note that a common practice to improve the performance of CNN-based SISR models is either increasing the depth of the network or deploying more complex architectures \cite{tai2017memnet, zhang2018image}. For instance, VDSR~\cite{kim2016accurate} is a 20-layer deep super-resolution convolutional network (VDSR), and more recent DRRN \cite{tai2017image}, SRDenseNet \cite{tong2017image}, and MemNet \cite{tai2017memnet} SISR models contain 52, 68, and 80 layers, respectively. However, deploying very deep CNN models for SISR comes with many obvious drawbacks such as difficult model training due to the gradient vanishing problem, slow running time, and a large number of model parameters \cite{ahn2018fast, lai2018fast, he2019mrfn}. In this paper, our motivation is to explore alternative techniques to improve the performance of SISR in terms of both accuracy and computational load. More specifically, we look into (1) designing better-performing feature extraction modules and (2) exploring more effective schemes for multiple feature fusion.

The first and most important improvement is incorporating an orientation-aware mechanism to the feature extraction modules in CNN-based SISR models. Our key observation is that image structures/textures are the complex combinations of features extracted in different directions (e.g., horizontal, vertical, and diagonal). Thus the optimal way of reconstructing missing image details should also be orientation-dependent. However, the existing CNN-based SISR models (e.g., SRCNN \cite{dong2014learning}, VDSR \cite{kim2016accurate}, DRRN \cite{tai2017image}, SRDenseNet \cite{tong2017image}, and MemNet \cite{tai2017memnet}) typically utilize standard $3\times3$ or $5\times5$ convolutional kernels, which are square-shaped and orientation-independent, to extract feature maps for the following super-resolution reconstruction. One possible solution to achieve orientation-aware SISR is to deploy convolutional kernels of various shapes in a single feature extraction module. In this paper, we proposed a novel Orientation-Aware feature extraction and fusion Module (OAM), which contains the mixture of 1D and 2D convolutional kernels (i.e., $5\times1$, $1\times5$, and $3\times3$) for extracting orientation-aware features.
 
The second improvement is to optimize the fusion scheme for integrating multiple features extracted in different directions and at various convolutional stages. Inspired by the channel attention mechanism for re-calibrating channel-wise features \cite{hu2018squeeze}, we firstly propose to incorporate local channel attention (LCA) mechanism within each orientation-aware feature extraction module. It performs the scene-specific fusion of multiple outputs of orientation-dependent convolutional kernels (e.g., horizontal, vertical, and diagonal) to generate more distinctive features for SISR. Moreover, we utilize global channel attention (GCA) mechanism as an effective technique to adaptively fuse low-level and high-level features extracted in hierarchically stacked OAMs. Finally, we experimentally evaluate a number of design options to identify the optimal way to utilize the CA mechanism for re-calculating channel-wise weights for the concatenated orientation-aware and hierarchical features.

Based on the above improvements (\textbf{a.} the orientation-aware feature extraction and \textbf{b.} the channel attention-based feature fusion), we present a compact but powerful CNN-based model for high-quality SISR via Channel Attention-based fusion of Orientation-Aware features (SISR-CA-OA). The proposed SISR-CA-OA model shows superior performance over the state-of-the-art SISR methods on multiple benchmark datasets, achieving more accurate image restoration results and faster running speed. Overall, the contributions of this paper are mainly summarized as follows:

\begin{itemize}
    \item We present a novel feature extraction module (OAM) containing a number of well-designed 1D and 2D convolutional kernels ($5\times1$, $1\times5$, and $3\times3$) to extract orientation-aware features.
    \item We design channel attention-based fusion schemes (LCA and GCA), which can adaptively combine features extracted in different directions and in hierarchically stacked convolutional stages.
    \item We present a powerful SISR-CA-OA model for high-quality SISR, achieving higher accuracy and faster running time compared with state-of-the-art deep learning-based SISR approaches \cite{dong2016image, kim2016accurate, kim2016deeply, tai2017image, tai2017memnet, hui2018two, li2018multi, haris2018deep, lim2017enhanced, zhang2018residual}.      
\end{itemize}    
	
In this paper, we make the following substantial extensions to our preliminary research work \cite{du2019orientation}. Firstly, we perform ablation studies to systematically validate the effectiveness of the proposed orientation-aware feature extraction technique. Secondly, we utilize the channel attention mechanism for the local fusion of features extracted in different directions and the global fusion of features extracted in hierarchical stages. Moreover, we investigate a number of design options to identify the optimal way to utilize the channel attention mechanism for feature fusion in SISR tasks. Thirdly, we significantly extend the experiments, comparing the proposed SISR-CA-OA model with a number of recently published SISR methods \cite{dong2016image, kim2016accurate, kim2016deeply, lai2018fast, tai2017image, tai2017memnet, hui2018two, li2018multi, haris2018deep, lim2017enhanced, zhang2018residual} using various benchmark datasets (Set5 \cite{bevilacqua2012low}, Set14 \cite{zeyde2010single}, B100 \cite{martin2001database}, Urban100 \cite{huang2015single}, and Manga109 \cite{fujimoto2016manga109}.).

The remainder of this paper is organized as follows. We first review a number of learning-based SISR models and different feature extraction/fusion techniques in Sec.~\ref{Related Work}. Then Sec.~\ref{Approach} provides details of important components in our proposed SISR-CA-OA model. Qualitative and quantitative evaluation results are provided in Sec.~\ref{experiments}, showing the superiority of our proposed method. Finally, Sec.~\ref{conclusion} concludes this paper.
 
\begin{figure*}[ht]
	\begin{center}
		\includegraphics[width=1\linewidth]{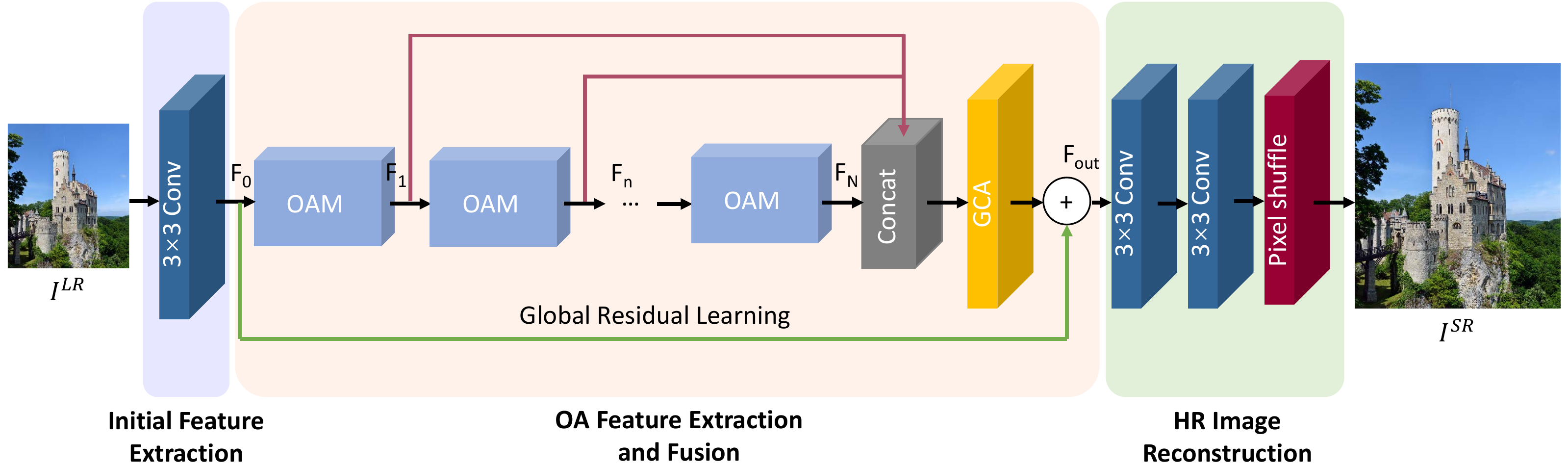}
	\end{center}
    \caption{The overall architecture of our proposed SISR-CA-OA model. Given the LR input image $I^{LR}$, we first employ a $3\times 3$ convolutional layer to extract low-level features. Then a number of OAMs are hierarchically stacked to infer the non-linear LR-to-HR mapping function. Note here we incorporate a local channel attention (LCA) mechanism for the local fusion of features extracted in different directions and a global channel attention (GCA) mechanism for the global fusion of features extracted in different convolutional stages. Global residual learning is added to ease the training process. Finally, we use two $3\times 3$ convolution layers and the pixel-shuffle operation to reconstruct the final HR image output $I^{SR}$.}
	\label{fig:OANet}
\end{figure*}

\section{Related Work}
\label{Related Work}

Over the past decades, developing effective SISR techniques to reconstruct a HR image from its corresponding single LR version has attracted extensive attention from both the academic fields and the industrial communities. In this work, we mainly focus on reviewing the existing CNN-based SISR methods which deploy various network architectures to construct distinctive feature representations for high-accuracy image restoration.

\subsection{Deep-learning-based SISR}
\label{Deep-learning-based Single Image Super-Resolution}
Dong et al. formulated the first 3-layer convolutional neural network model (SRCNN) to implicitly learn the end-to-end mapping function between the LR and HR images \cite{dong2014learning, dong2016image}. Following this pioneering work, Kim et al. presented deeper networks (VDSR \cite{kim2016accurate} and DRCN \cite{kim2016deeply}) to generate more distinctive feature over larger image regions for more accurate image restoration. To alleviate the gradient vanishing problem that occurs when training a deep CNN model, they integrate a global residual learning architecture, which is firstly proposed by He et al. \cite{he2016deep} into their SISR models. Dong et al. proposed to deploy a deconvolution layer to up-scale the feature maps at the end of the neural network to achieve faster speed and better reconstruction accuracy \cite{dong2016accelerating}. For the same purpose, Shi et al. proposed a pixel-shuffle operation for fast and accurate upscaling of the LR images via rearranging the feature maps \cite{shi2016real}. 

It is noted that the most previously proposed SISR methods attempted to achieve more accurate restoration results by either increasing the depth of the network or deploying more complex architectures. For instance, Tai et al. developed a 52-layer DRRN model \cite{tai2017image} which deploys local and global residual learning and recursive layers and a 84-layer MemNet model \cite{tai2017memnet} which contains persistent memory units and multiple supervisions. More recently, some very deep CNN models such as RDN \cite{zhang2018residual}, D-DBPN \cite{haris2018deep}, MSRN \cite{li2018multi}, and RCAN \cite{zhang2018image} are trained using the high-resolution DIV2K \cite{DIV2Kdataset} dataset (containing 800 training images of 2K resolution), achieving the state-of-the-art SISR performance. However, their training process took a long time to complete and cannot deliver real-time processing speed. In this paper, we aim to develop better-performing feature extraction modules and more effective feature fusion schemes to improve the performance of SISR in terms of both accuracy and computational load.  

\subsection{Feature Extraction and Fusion}

The existing CNN-based SISR models such as SRCNN \cite{dong2016image}, VDSR \cite{kim2016accurate}, DRRN \cite{tai2017image}, SRDenseNet \cite{tong2017image}, and MemNet \cite{tai2017memnet} typically deploy square-shaped $3\times3$ or $5\times5$ convolutional kernels to extract feature maps for the following super-resolution reconstruction. Li et al. proposed to utilize convolution kernels of different sizes to construct scale-dependent image features for better restoration of both large-size structures and small-size details \cite {li2018multi}. He et al. designed multi-receptive-model to extract features in different receptive fields from local to global \cite{he2019mrfn}. In some other computer vision tasks, researchers attempted to deploy a number of kernels of different shapes to generate more comprehensive and distinctive features. For instance, Liao et al. introduced TextBoxes \cite{liao2017textboxes} which employed irregular $1\times5$ convolutional filters to yield rectangular receptive fields for text detection. In Google Inception-Net V2 \cite{ioffe2015batch}, Ioffe et al. utilized different $1 \times n$ and $n \times 1$ rectangular kernels instead of $n \times n$ square kernels so as to decrease parameters. Li et al. introduced multi-scale feature extraction blocks which contain convolutional kernels of various shapes (e.g., $3\times3$, $1\times5$, $5\times1$, $1\times7$, $7\times1$, and $1\times1$ ) to generate informative features for classification of eye defects \cite{li2019large}. In this paper, we present a novel feature extraction module which contains the mixture of 1D and 2D convolutional kernels (i.e., $5\times1$, $1\times5$, and $3\times3$) for computing orientation-aware features for the SISR task.

To perform high-accuracy HR image reconstruction, it is important to utilize the hierarchical features extracted in different convolutional stages. Many deep CNN models added dense skip connections to combine low-leave features extracted in shallower layers with semantic features computed in deeper layers to generate more informative feature maps and tackle the problem of gradient vanishing  \cite{kim2016accurate,tai2017image,lim2017enhanced, huang2017densely}. Huang et al. introduced dense skip connections into DenseNet models, reusing the feature maps of preceding layers to enhance the representation of features, and alleviate the problem of gradient vanishing \cite{huang2017densely}. Zhang et al. proposed to fuse hierarchical feature maps extracted in stacked residual dense blocks, achieving better reconstruction results \cite{zhang2018residual}. Tai et al. proposed densely concatenated memory blocks to reconstruct accurate details for the task of image restoration \cite{tai2017memnet}. Given the channel-wise concatenated features, it is desirable to design an effective fusion scheme for selecting the most distinctive ones. The channel attention (CA) mechanism, which was initially proposed for image classification tasks \cite{xiao2015application,wang2017residual}, has recently been adopted to solve the challenging SISR problem via re-calibrating the feature responses towards the most informative and important channels of the feature maps \cite{zhang2018image, hu2019channel, CA2018BMVC}.  In this paper, we design/optimize CA-based fusion schemes to adaptively combine features extracted in different directions and in hierarchically stacked convolutional stages.

\section{Approach}
\label{Approach}

In this section, we propose a CNN-based model for fast and accurate SISR via Channel Attention-based fusion of Orientation-Aware features (SISR-CA-OA). We first present the architecture of the proposed SISR-CA-OA model. Then we provide details of the key building blocks of the SISR-CA-OA model include \textbf{a.} the orientation-aware feature extraction modules and \textbf{b.} the channel attention-based multiple feature fusion schemes. 

\subsection{Network Architecture}
\label{Network Structure}
As illustrated in Fig.~\ref{fig:OANet}, the SISR-CA-OA model consists of three major processing steps: (1) initial feature extraction on the input LR image $I^{LR}$, (2) orientation-aware feature extraction and fusion, (3) HR image reconstruction. 

Given a LR input image $I_{LR}$ ($H \times W$), a $3\times3$ convolutional layer is firstly deployed to extract low-level features $F_{0} \in \mathbb{R}^{C\times H\times W}$ ($C$- channel number, $H$ - image height, $W$ - image width) as
\begin{equation}
	F_{0} = Conv_{3\times3}(I^{LR}),
\end{equation}
where $Conv_{3\times3}$ denotes the convolution operation using a $3\times 3$ kernel. Then, the extracted $F_{0}$ is fed to a number of stacked OAMs to compute orientation-aware features $F_{n}$ in different convolutional stages (more details of orientation-aware feature extraction are provided in Sec.~\ref{Oreientation-aware FET}). Within each OAM, we design a LCA-based fusion scheme to perform the scene-specific fusion of multiple orientation-aware features. Moreover, we present a feature fusion technique based on the GCA mechanism to integrate low-level and high-level semantic features extracted in hierarchically stacked OAMs. Detail information of these channel attention-based feature fusion schemes is provided in Sec.~\ref{CA}.

We adopt the global residual learning technique in the proposed SISR-CA-OA model by adding an identity branch from its initial input $F_0$ to the hierarchically fused feature $F^{GCA}$ ({\color{green}green} line in Fig.~\ref{fig:OANet}) as
\begin{equation}
	F_{out} = F_{0} + F^{GCA},
\end{equation}
where $+$ calculates the sum of feature maps $F_{0}$ and $F^{GCA}$ at the same spatial locations and channels. The computed feature maps $F_{out}$ is then fed to two convolutional layers and an up-sampling layer to reconstruct the HR image. For a $\times R$ upscaling SISR task, two $3\times3$ convolutional layers are utilized to convert the channel number of $F_{out}$ from $C$ to $R \times R$ and the up-sampling layer performs the pixel shuffle operation \cite{shi2016real} to reconstruct the super-resolved output $I^{SR}$ ($RH \times RW$).

The SISR-CA-OA model is optimized by minimizing the pixel-wise difference between the predicted super-resolved image $I^{SR}$ and corresponding ground truth $I^{GT}$. In this paper, the training and testing of SISR models are performed on the Y channel (i.e., luminance) of transformed YCbCr space \cite{dong2014learning, tai2017image, tai2017memnet, zhang2018residual}. We adopt the two-parameter weighted Huber loss function to drive the weights learning \cite{he2019mrfn}. The weighted Huber loss function sets larger back-propagated derivatives to accelerate the training process when the training residuals are significant. In comparison, it linearly decreases the back-propagated derivative to zero when the residual value is approaching zero. As a result, the weighted Huber loss combines the advantages of $L1$ and $L2$ loss functions and fits the reconstruction error more effectively.

\subsection{Orientation-aware Feature Extraction}
\label{Oreientation-aware FET}

The CNN-based SISR models typically deploy a set of convolutional kernels to extract semantic features for HR image reconstruction. The $3\times3$ convolutional kernel is the most widely used option in many state-of-the-art SISR models such as VDSR \cite{kim2016accurate}, DRCN \cite{kim2016deeply}, DRRN \cite{tai2017image}, SRDenseNet \cite{tong2017image}, EDSR \cite{lim2017enhanced}, and TSCN \cite{hui2018two}. More recently, some researchers adopted convolutional kernels of larger sizes (e.g., $5\times5$ or $7\times7$) to generate multi-scale features \cite{li2018multi,CMSCN2018,he2019mrfn}. It is noted that the existing SISR models typically utilize square-shaped and orientation-independent convolutional kernels (e.g., $3\times3$ or $5\times5$) to extract feature maps for reconstructing image structures/textures in different directions. One possible solution to build more distinctive features for high-accuracy SISR is incorporating multiple convolutional kernels of various shapes for extracting orientation-aware features in a single feature extraction module.

\begin{figure}[ht]
	\begin{center}
		\includegraphics[width=0.99\linewidth]{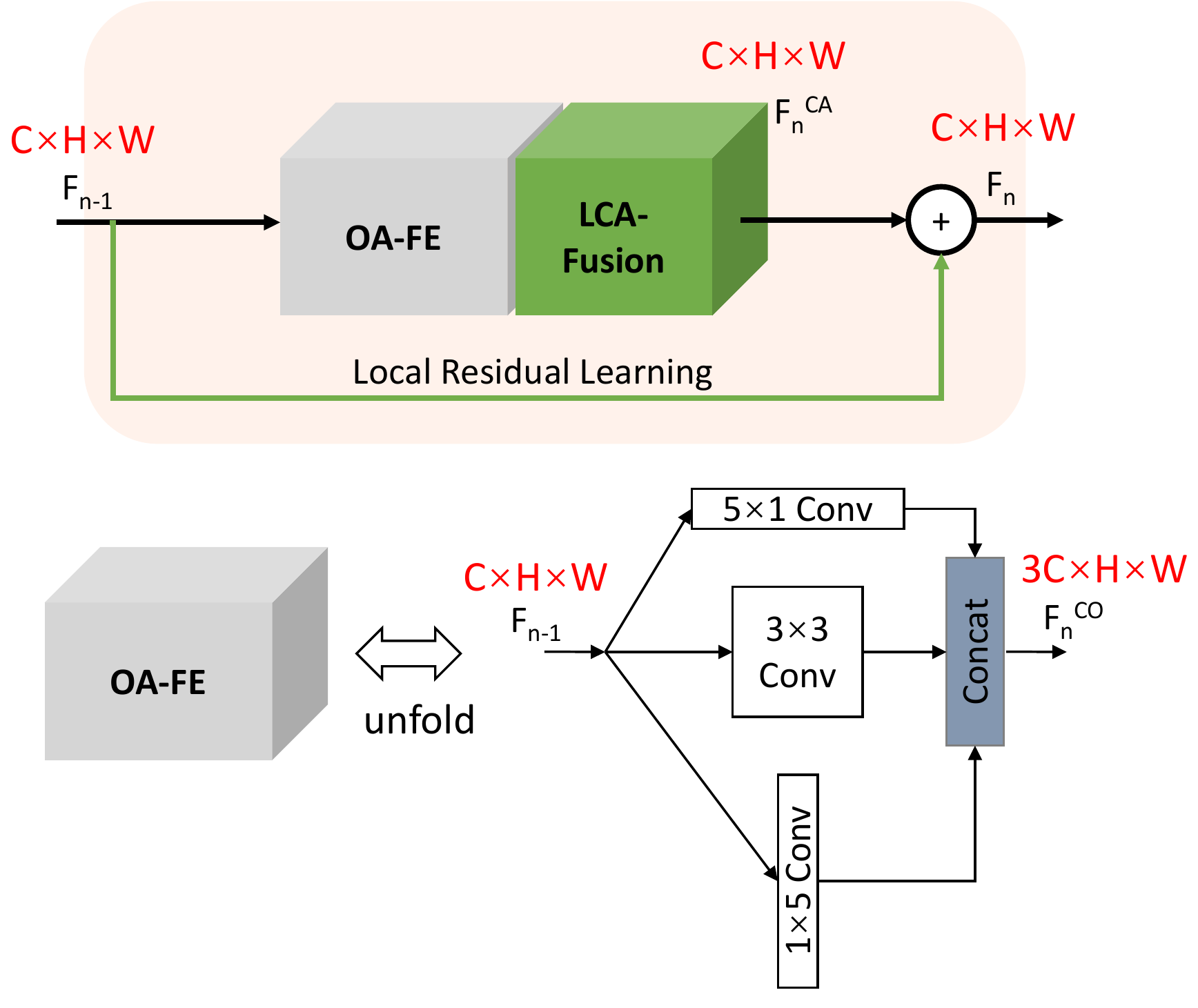}
	\end{center}
    \caption{The architecture of the backbone OAM for feature extraction. The input is firstly fed into three individual convolutional layers using kernels of different shapes ($3\times 3$, $1\times 5$, and $5\times 1$). Then the extracted orientation-aware feature maps are concatenated and then processed by a LCA-based fusion block  to adaptively re-calibrate the channel-wise weights towards the most informative and important channels. The residual learning is also added to OAM to ease the training process.}
	\label{fig:OAM}
\end{figure}

In each OAM, we deploy a standard convolutional layer using the $3\times 3$ square-shaped kernel and two additional convolutional layers using 1D kernels (i.e., $1\times 5$ and $5\times 1$) to extract features in different directions, as  illustrated in Fig.~\ref{fig:OAM}. Let $F_{n-1}$ denote the input feature maps of $n$-th OAM, the orientation-aware feature maps $F_{n}^{H}$, $F_{n}^{V}$, and $F_{n}^{D}$ are computed as:
\begin{equation}
	F_{n}^{H} = Conv_{5\times1}(F_{n-1}),
\end{equation}
\begin{equation}
	F_{n}^{V} = Conv_{1\times5}(F_{n-1}),
\end{equation}
\begin{equation}
	F_{n}^{D} = Conv_{3\times3}(F_{n-1}),
\end{equation}
where $Conv_{5\times1}$, $Conv_{1\times5}$ and $Conv_{3\times3}$ represent the convolution operations using the $5\times 1$, $1\times 5$, and $3\times 3$ kernels, respectively. The 1D $1\times 5$ kernel only considers the information in the horizontal direction thus can better extract vertical features. On the other hand, the $5\times 1$ kernel only covers vertical pixels thus is more suitable for the extraction of horizontal features. In this manner, we propose to firstly extract orientation-aware features in different directions (e.g.,  horizontal, vertical, and diagonal) and then perform scene-specific fusion to generate more informative features for SISR. In Sec.~\ref{analysis-FET}, we will set up experiments to systematically evaluate the effectiveness of the proposed orientation-aware feature extraction technique.

\subsection{Channel Attention-based Feature Fusion}
\label{CA}
\begin{figure}[ht]
	\begin{center}
	\begin{minipage}[t]{0.99\linewidth}
		\includegraphics[width=1\linewidth]{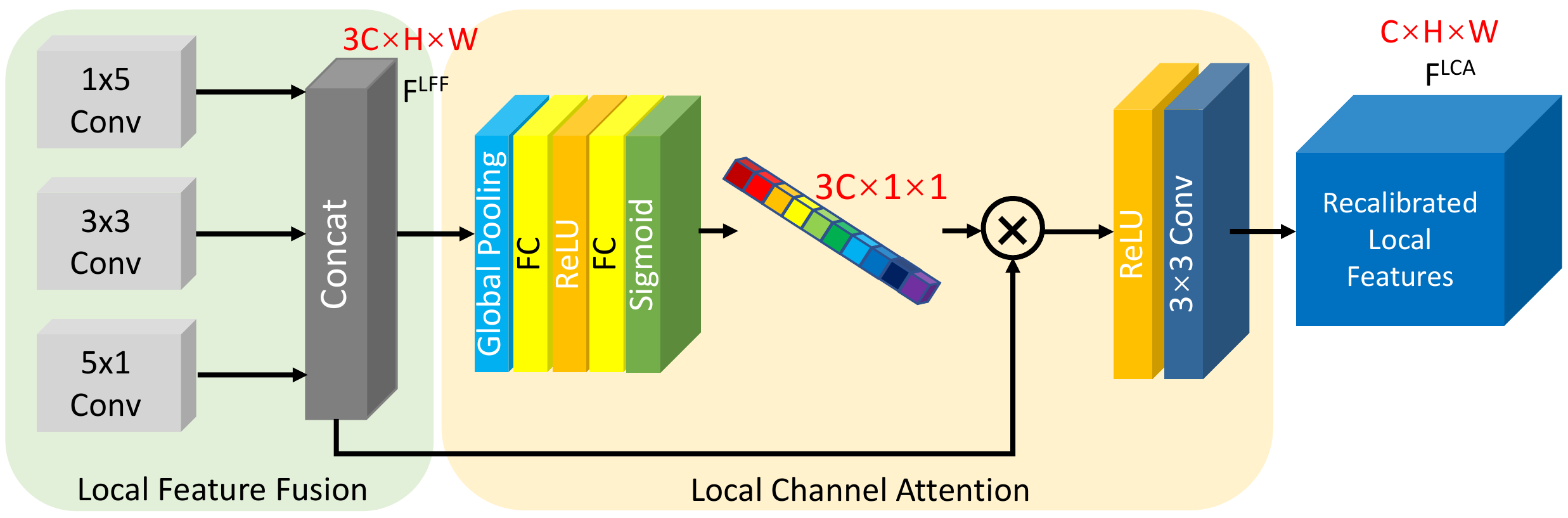}
		\centering{(a) LCA-based fusion}
	\end{minipage}
	\hspace{3mm}
		\begin{minipage}[t]{0.99\linewidth}
		\includegraphics[width=1\linewidth]{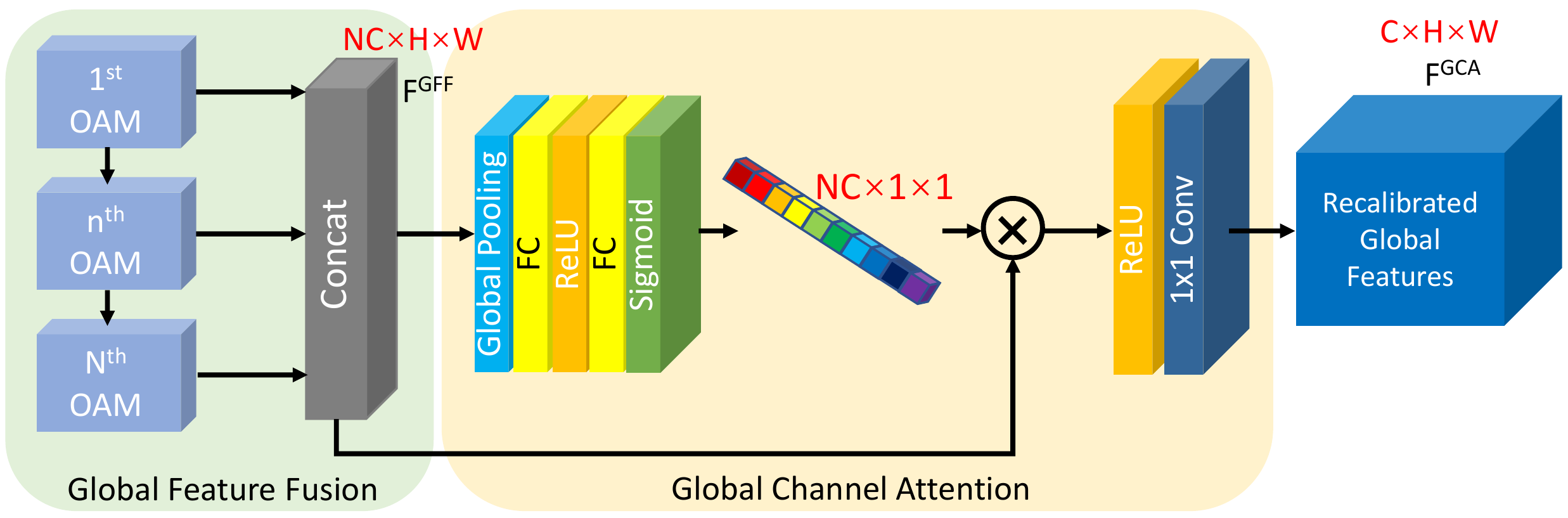}
		\centering{(b) GCA-based fusion}
	\end{minipage}
		\end{center}
    \caption{CA-based fusion schemes for integrating multiple features extracted in different directions and different convolutional stages. (a) The LCA mechanism and (b) the GCA mechanism for the channel-wise weights re-calibration of concatenated orientation-aware and hierarchical features.}
	\label{fig:LCA&GCA}
\end{figure}

Previous research works have proven the effectiveness of Channel attention (CA) mechanism \cite{hu2018squeeze}, providing a way to re-calibrate channel-wise features via explicitly modeling interdependencies between channels in the task of super-resolution \cite{lu2018channel,cheng2018sesr,zhang2018image, CA2018BMVC}. In this paper, we design CA-based fusion schemes to integrate multiple features extracted in different directions and different convolutional stages. More specifically, we incorporate a LCA mechanism within each OAM, performing the scene-specific fusion of orientation-aware features, as shown in Fig.~\ref{fig:LCA&GCA} (a). Moreover, we present a GCA-based fusion scheme to integrate low-level and high-level features extracted in various convolutional stages, as illustrated in Fig.~\ref{fig:LCA&GCA} (b). 

\subsubsection{Fusion of Orientation-aware Features}
\label{LCA}

Within the $n$-th OAM, the computed orientation-aware feature maps $F_{n}^{H}$, $F_{n}^{V}$, and $F_{n}^{D}$ are firstly combined through a simple concatenation operation as 
\begin{equation}
	F_{n}^{CO} = [F_{n}^{H}, F_{n}^{V}, F_{n}^{D}],
\end{equation}
where $[\cdot]$ denotes the concatenation operation, and $F_{n}^{CO} \in \mathbb{R}^{3C\times H\times W}$ is the concatenated orientation-aware features. Given features extracted in different directions, we deploy the LCA mechanism to emphasize the informative features as well as to suppress redundant ones, performing an adaptive fusion of orientation-aware features. As illustrated in Fig.~\ref{fig:LCA&GCA} (a), LCA firstly shrinks the concatenated orientation-aware features $F_{n}^{CO} \in \mathbb{R}^{3C\times H\times W}$ along the spatial dimensions $H\times W$ through a global average pooling operation. A channel-wise descriptor $z \in \mathbb{R}^{3C\times 1\times 1}$ is computed and the $c$-th element of $z$ is
\begin{equation}
	z^{CO}_c = GP(F_{n,c}^{CO}) = \dfrac{1}{H \times W} \sum_{h=1}^{H} \sum_{w=1}^{W} F_{n,c}^{CO}(h,w),
	\label{Eq10}
\end{equation}
where $GP(\cdot)$ denotes the global average pooling operation and $F_{n,c}^{CO}(h,w)$ is the value at coordinate position $(h,w)$ of the $c$-th channel of $F_{n}^{CO}$. A gating mechanism \cite{hu2018squeeze} consisting of two fully connected (FC) layers and a ReLU activation function is then deployed to assign weights to different feature channels as
\begin{equation}
	\alpha^{CO} = \sigma(FC(\delta(FC(z^{CO})))),
	\label{Eq11}
\end{equation}
where $FC(\cdot)$ are the FC layers, $\delta(\cdot)$ represents the ReLU function. Note a sigmoid function $\sigma(\cdot)$ is utilized to adjust the channel attention weights to the range between 0 and 1. The first FC layer reduces the channel dimension to $\frac{1}{s}$ and the second FC layer increases the channel dimension from $\frac{3C}{s}$ back to $3C$. The re-calculated output $F_{n}^{CO-LCA}$ is obtained by rescaling the concatenated orientation-aware $F_{n}^{CO}$ with the attention weights $\alpha^{CO}$ channel-wisely. More specifically, the $c$-th channel of $F_{n}^{CO-LCA}$ can be calculated as 
\begin{equation}
	F_{n,c}^{CO-LCA} =  \alpha^{CO}_{c} \cdot F_{n,c}^{CO}.
	\label{Eq12}
\end{equation}
It is worth mentioning that the scene-specific channel weights $ \alpha^{CO}_{c}$ are completely self-learned without supervision. As a result, LCA adaptively assigns higher weights for the informative features as well as to suppress redundant ones to perform the adaptive fusion of orientation-aware features.

The computed features $F_{n}^{CO-LCA}$ is then activated using a ReLU function and fed into a $3\times3$ convolutional layer, squeezing the channel number of $F_{n}^{CO-LCA}$ from $3C$ to $C$ as
\begin{equation}
	F_{n}^{LCA} = Conv_{3\times3}(\delta(F_{n}^{CO-LCA})).
\end{equation}
Local residual learning technique ({\color{green}green} line in Fig.~\ref{fig:OAM}~(a)) is also deployed to alleviate the gradient vanishing/exploring problem \cite{zhang2018residual, lim2017enhanced}, thus the output of the $n$-th OAM is
\begin{equation}
	F_{n} = F_{n-1} + F_{n}^{LCA}.
\label{Eq14}
\end{equation}

\subsubsection{Fusion of Hierarchical Features}
\label{GCA}

It is important to utilize the hierarchical features extracted in different convolutional stages for high-accuracy SISR \cite{zhang2018residual,wang2018deep,tong2017image,kim2016deeply,lee2015deeply}. As illustrated in Fig.~\ref{fig:OAM} (b), we deploy a GCA mechanism to re-calibrate the channel weights for the concatenated hierarchical features, adaptively combining semantic features extracted in deeper layers and low-level features extracted in shallower layers. Given the outputs of $N$ OAMs ($F_1,F_2,\cdots,F_N$), we compute the concatenated hierarchical features $F^{CH} \in \mathbb{R}^{NC\times H \times W}$ as 
\begin{equation}
	F^{CH} = [F_1,F_2,\cdots,F_N].
\end{equation}
Similarly, the GCA mechanism calculates channel-wise attention weights for the concatenated hierarchical features $F^{CH}$ as
\begin{equation}
	\alpha^{CH} = \sigma(FC(\delta(FC(GP(F^{CH}))))).
\end{equation}
The re-calibrated concatenated hierarchical features $F^{CH-GCA}$ are calculated by rescaling $F^{CH}$ with the attention weights $\alpha^{CH} $ channel-wisely as
\begin{equation}
	F_{c}^{CH-GCA} =  \alpha^{CH}_{c} \cdot F_{c}^{CH}.
	\label{Eq112}
\end{equation}
Note the computed $F^{CH-GCA}$ is also activated using a ReLU function to embed more nonlinear terms into the network. Moreover, a $1\times 1$ convolutional layer is utilized to compress the channel number from $N\times C$ to $C$. The final output of GCA-based hierarchical feature fusion is  
\begin{equation}
	F^{GCA} = Conv_{1\times1}(\delta(F^{CH-GCA})).
\label{Eq13}
\end{equation}

In Sec.~\ref{analysis-LCA}, we set up systematical experiments to validate the effectiveness of the proposed LCA/GCA-based fusion schemes. Moreover, we investigate a number of design options for integrating the CA mechanism in our proposed SISR-CA-OA model to achieve better fusion of multiple features extracted in different orientations and different convolutional stages. 

\section{Experimental Results}
\label{experiments}
In this section, we systematically evaluate the performance of our proposed SISR-CA-OA model and compare it with the state-of-the-art SISR methods quantitatively and qualitatively on a number of commonly used benchmark datasets. 

\subsection{Datasets and Metrics}
\label{datasets}
\textbf{Training:} Following \cite{kim2016accurate,kim2016deeply}, we train a light-weight version of SISR-CA-OA model consisting of 10 OAMs on RGB91 dataset from Yang et al. \cite{yang2010image} and another 200 images from Berkeley Segmentation Dataset (BSD) \cite{martin2001database}. Moreover, we make use of the DIVerse 2K resolution image dataset (i.e., DIV2K) \cite{DIV2Kdataset} to train an enhanced SISR-CA-OA model (SISR-CA-OA$\ast$) which contains 64 stacked OAMs.  Three commonly used data augmentation techniques are utilized to expand our training dataset, including 1. Rotation: rotate image by $90^\circ$, $180^\circ$ and $270^\circ$. 2. Flipping: horizontally flip image. 3. Scaling: downscale image with the scale factors of 0.9, 0.8, 0.7, 0.6 and 0.5. After the augmentation, we randomly crop these images into a number of sub-images ($48\times48$ for the RGB91 and BSD datasets and $96\times96$ for the DIV2K dataset). The LR images are obtained by down-sampling corresponding HR images using bicubic interpolation. 

\textbf{Testing:} Five commonly used public benchmark datasets are utilized for evaluating the performance of our SISR-CA-OA model. Set5 \cite{bevilacqua2012low} and Set14 \cite{zeyde2010single} are widely used datasets in SISR tasks. B100 \cite{martin2001database} contains 100 natural images collected from BSD, and Urban100 \cite{huang2015single} consists of 100 real-world images which are rich of structures. Manga109 dataset, which consists of a variety of 109 Japanese comic books, is also employed \cite{fujimoto2016manga109}.

\textbf{Evaluation Metrics:} Peak signal-to-noise-ratio (PSNR) and structural similarity index (SSIM) \cite{Wang2004SSIM} are used for SISR performance evaluation. The training and testing of SISR models are performed on the Y channel (i.e., luminance) of transformed YCbCr space \cite{dong2014learning, tai2017image, tai2017memnet, zhang2018residual}. For a fair comparison, we crop pixels near image boundary according to \cite{dong2016image}. 

\subsection{Implementation Details}
\label{Implementation Details}

We implement our SISR-CA-OA model with Caffe\cite{jia2014caffe} platform and train this model by optimizing Modified Huber Loss function on a single NVIDIA Quadro P6000 GPU with Cuda 9.0 and Cudnn 7.1 for 60 epochs. When training our model, we only consider the luminance channel (Y channel of YCbCr color space) in our experiments. Adam\cite{kingma2014adam} solver is utilized to optimize the weights by setting $\beta_{1}=0.9$, $\beta_{2}=0.999$ and $\varepsilon=1e^{-8}$. In each training batch, we randomly crop these augmented training images into $48\times 48$ patches and the batch size is set to 64 for training our SISR-CA-OA. The initial learning rate is set to $1e^{-4}$ and halved after 50 epochs. Training of SISR-CA-OA models approximately takes two days. When training our SISR-CA-OA for scale factors $\times3$ and $\times4$, we initialize the weights with pre-trained $\times2$ model and decrease the learning rate to $1e^{-5}$. The source codes will be made publicly available in the future.

\subsection{Performance Analysis}
\label{Model Analysis}

In this section, we set up ablation experiments to evaluate the effectiveness of (1) orientation-aware feature extraction, and (2) channel attention based feature fusion.

\subsubsection{Orientation-aware Feature Extraction}
\label{analysis-FET} 

We evaluate the performance of three different designs of residual blocks including (a) a standard residual block which consists of two $3\times3$ convolutional layers and a ReLU activation layer \cite{lim2017enhanced, CA2018BMVC}, (b) a residual block which utilizes three individual square-shaped ($3\times3$) convolutional kernels to extract features, (c) our propose orientation-aware residual block which uses a standard $3\times 3$ and two additional $1\times 5$ and $5\times 1$ convolutional kernels to extract features in different directions. For a fair comparison, three different residual blocks are implemented in the same EDSR the baseline model \cite{lim2017enhanced} without performing CA-based feature fusion. We set the number of residual blocks $N=10$ and the channel number of each convolutional layer to 64. Tab.~\ref{tab:tab1} summarizes the quantitative evaluation results (PSNR and SSIM) on Set5, Urban100, and Manga109 datasets with the scale factor $\times2$. First of all, it is experimentally demonstrated that incorporating multiple convolutional kernels within a residual block (i.e., design (b) and (c)) can generally construct more distinctive features and achieve higher SISR accuracy. Moreover, the residual block incorporating a mixture of 1D and 2D convolutional kernels ($3\times3$, $1\times 5$, and $5\times 1$) performs better than the one based on three square-shaped kernels ($3\times3$), achieving higher PSNR and SSIM indexes with fewer parameters. The experimental results manifest the effectiveness of the orientation-aware design, utilizing convolutional kernels of various shapes to extract orientation-aware features for more accurate reconstruction of image structures/textures in different directions.

\begin{figure}[ht]
	\begin{center}
		\begin{minipage}[t]{0.9\linewidth}
		\includegraphics[width=1\linewidth]{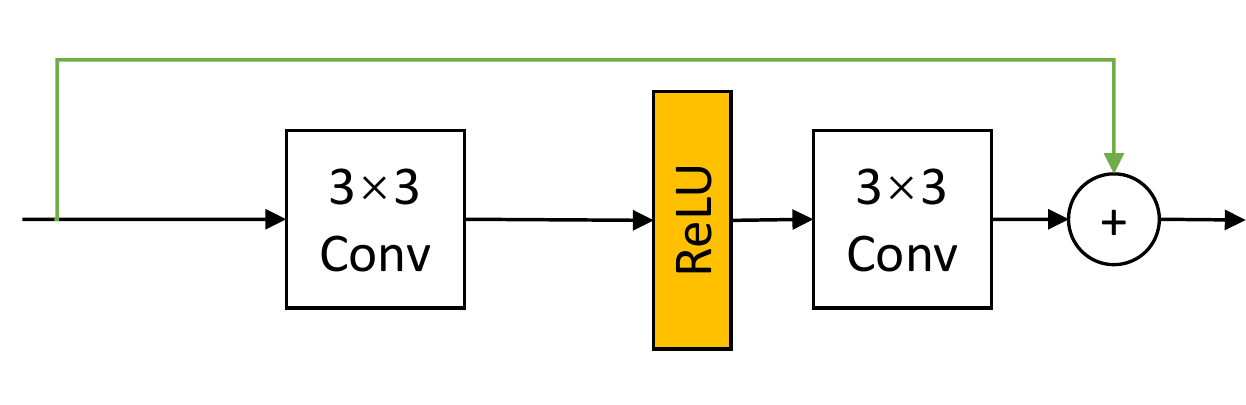}
		\centering{(a)}
	\end{minipage}
			\begin{minipage}[t]{0.9\linewidth}
		\includegraphics[width=1\linewidth]{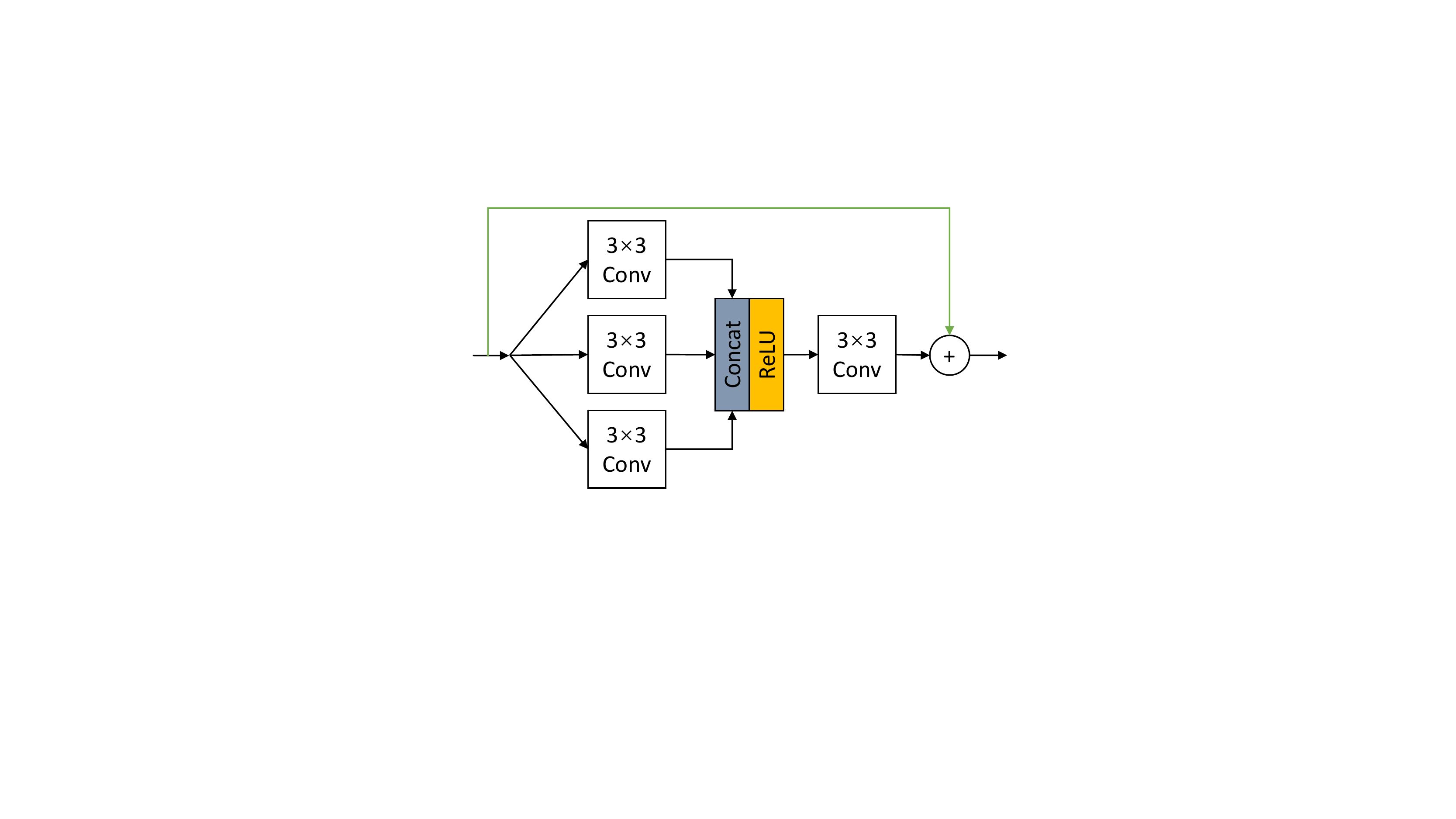}
		\centering{(b)}
			\end{minipage}
		\begin{minipage}[t]{0.9\linewidth}
		\includegraphics[width=1\linewidth]{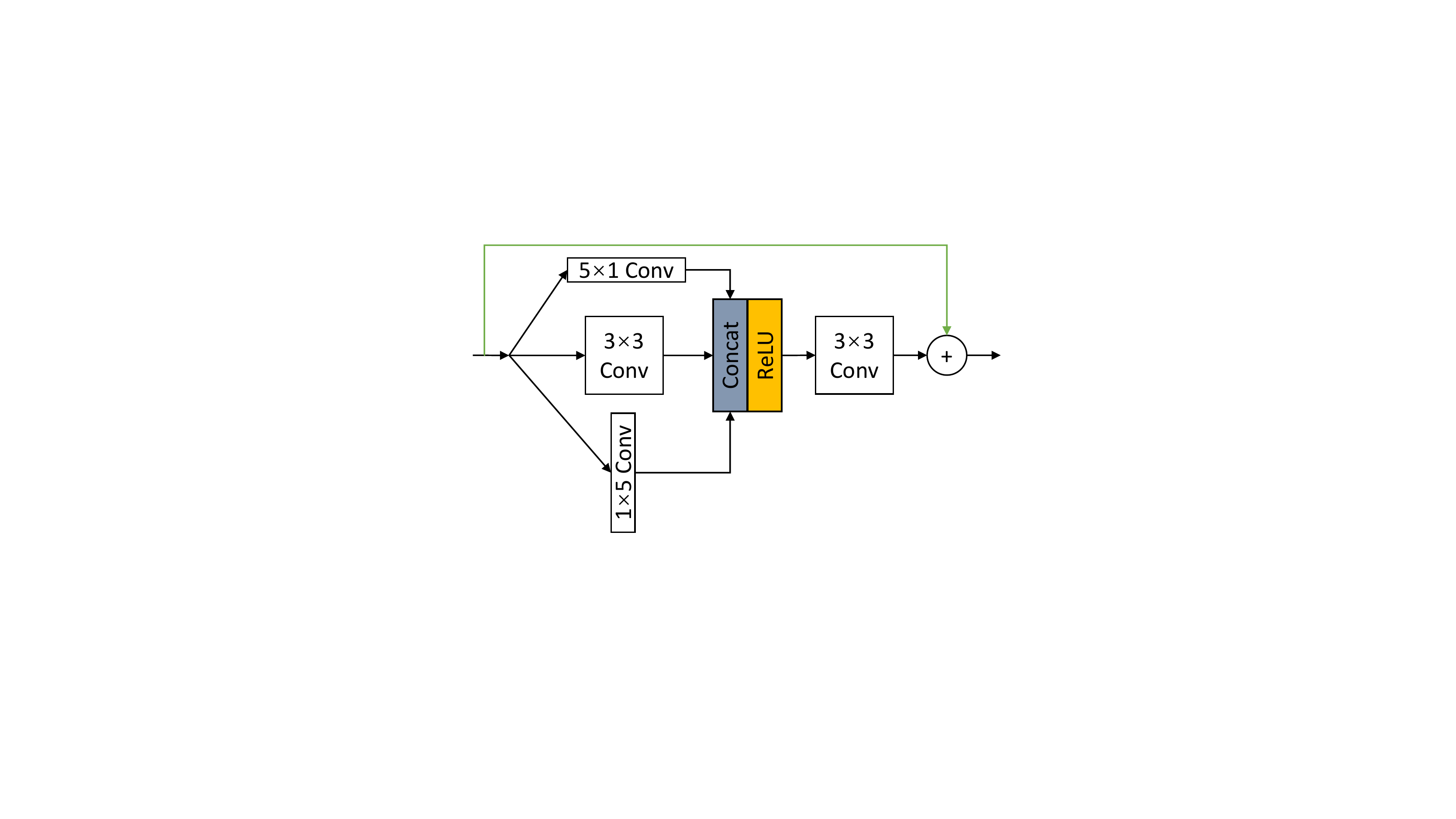}
		\centering{(c)}
	\end{minipage}
	\end{center}
	\caption{Structures of three residual block designs: (a) A residual block utilized in many SISR models \cite{lim2017enhanced,CA2018BMVC}, (b) A residual block which contains three square-shaped ($3\times3$) convolutional kernels for feature extraction, (c) Our proposed orientation-aware residual block which incorporates a mixture of 1D and 2D convolutional ($3\times3$, $1\times 5$, and $5\times 1$) kernels.}
	\label{fig:analysis1}
\end{figure}


\begin{table}[ht]
	\caption{Experimental evaluation of residual blocks of three different designs. PSNR(dB) and SSIM metrics are calculated on Set5, Urban100 and Manga109 datasets with scale factor $\times 2$.}
	\label{tab:tab1}
	\begin{center}
	\begin{tabular}{|cc||ccc|}
		\hline
		\multicolumn{2}{|c||}{\multirow{2}{*}{}} & \multicolumn{3}{c|}{Different Residual Blocks} \\ 
		\multicolumn{2}{|c||}{\multirow{2}{*}{}}            & Design (a)     & Design (b)     & Design (c)    \\ \hline \hline
		\multirow{2}{*}{Set5}       & PSNR   & 37.76          & 37.79          & \textbf{37.84}         \\ 
		                                         & SSIM      & 0.9596        & 0.9596         & \textbf{0.9600}        \\ \hline
		\multirow{2}{*}{Urban100}   & PSNR   & 31.17          & 31.35          & \textbf{31.41}        \\ 
		& SSIM      & 0.9187        & 0.9206         & \textbf{0.9216}       \\ \hline
		\multirow{2}{*}{Manga109}   & PSNR   & 37.89          & 37.95          & \textbf{38.02}        \\ 
		& SSIM      & 0.9747        & 0.9748         & \textbf{0.9749}       \\ \hline
	\end{tabular}
	\end{center}
\end{table}


\subsubsection{CA-based Feature Fusion}
\label{analysis-LCA}

\begin{figure*}[ht]
	\begin{center}
		\begin{minipage}[t]{.7\linewidth}
		\includegraphics[width=1\linewidth]{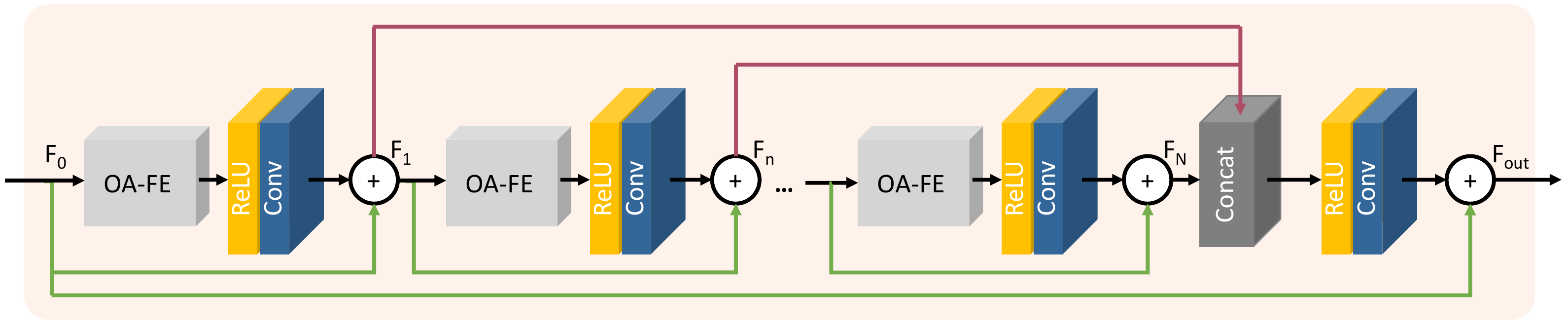}
		\centering{(a)}
	\end{minipage}\\
			\begin{minipage}[t]{.7\linewidth}
		\includegraphics[width=1\linewidth]{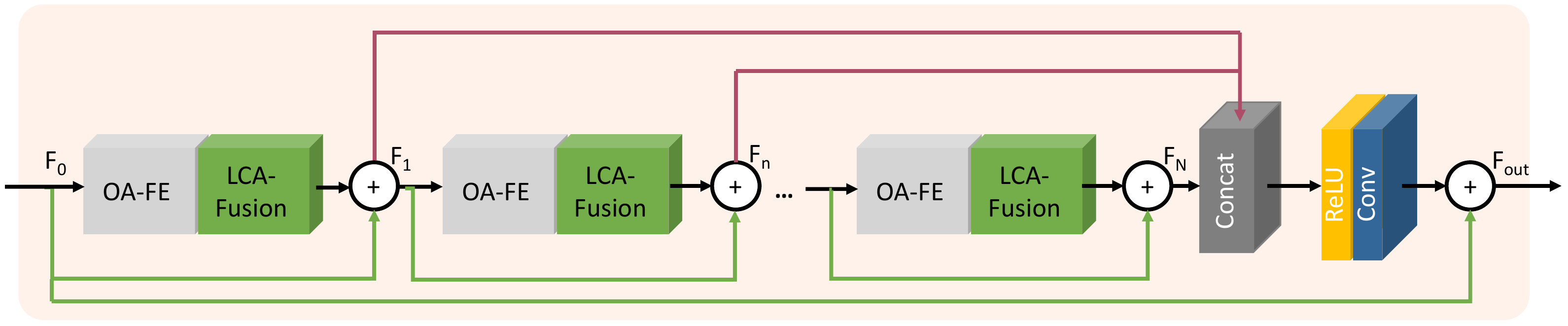}
		\centering{(b)}
	\end{minipage}\\
		\begin{minipage}[t]{.7\linewidth}
		\includegraphics[width=1\linewidth]{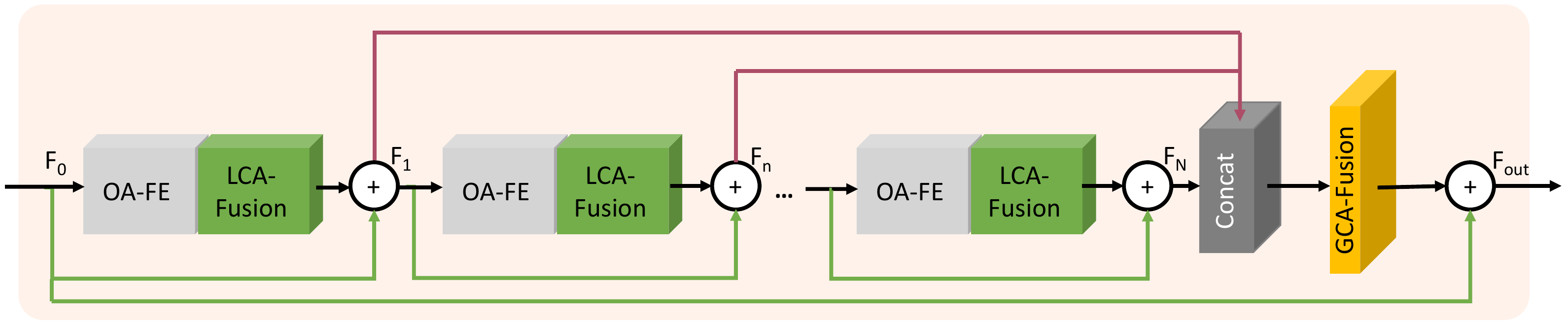}
		\centering{(c)}
	\end{minipage}
		\end{center}
	\caption{Ablation experiments to evaluate the effectiveness of the proposed CA-based feature fusion schemes. (a) Feature fusion without utilizing the LCA/GCA mechanisms for channel weights re-calibration, (b) Feature fusion with LCA-based channel weights re-calibration only, (c) Feature fusion incorporating both LCA and GCA mechanisms.}
	\label{fig:analysis2-3}
\end{figure*}

\begin{table}[ht]
    \caption{Comparative results of three ablation experiments with/without performing the CA-based channel weights re-calibration. PSNR(dB) and SSIM metrics are calculated on Set5, Urban100 and Manga109 datasets with scale factor $\times 2$.}
	\label{tab:tab9}
	\begin{center}
		\begin{tabular}{|cc||ccc|}
			\hline
			\multicolumn{2}{|c||}{\multirow{2}{*}{}}      & \multicolumn{3}{c|}{Different Fusion Schemes} \\ 
			\multicolumn{2}{|c||}{\multirow{2}{*}{}}               &     Exp. A     &     Exp. B    &     Exp. C    \\
			\hline \hline
			\multicolumn{2}{|c||}{LCA}            &   $\times$    & $\checkmark$ & $\checkmark$ \\ 
			\multicolumn{2}{|c||}{GCA}            &   $\times$    &   $\times$   & $\checkmark$ \\ \hline \hline
			\multirow{2}{*}{Set5}     & PSNR  &     37.86     &     37.90    & \textbf{37.97}\\ 
			& SSIM     &    0.9659    & 0.9600    & \textbf{0.9605}\\ \hline
			\multirow{2}{*}{Urban100} & PSNR  &     31.45     &     31.51    & \textbf{31.57}\\ 
			& SSIM     &    0.9217    & 0.9220     & \textbf{0.9226}\\ \hline
			\multirow{2}{*}{Manga109} & PSNR  &     38.03     &     38.11    & \textbf{38.38}\\ 
			& SSIM     &    0.9748    & 0.9750     & \textbf{0.9755}\\ \hline
		\end{tabular}
	\end{center}
\end{table}

In this section, we set up three ablation experiments to evaluate the effectiveness of the proposed CA-based feature fusion schemes, as illustrated in Fig.~\ref{fig:analysis2-3}. In Experiment A, the concatenated orientation-aware and hierarchical features are directly fed to a ReLU activation layer and a convolution layer to compute the fused feature map without utilizing the LCA/GCA mechanisms for channel weights re-calibration. In Experiment B, we only perform the LCA-based fusion in individual OAMs to combine multiple outputs of orientation-dependent convolutional kernels. In Experiment C, we perform both LCA-based orientation-aware feature fusion and GCA-based hierarchical feature fusion. Tab.~\ref{tab:tab9} shows the comparative results (PSNR and SSIM) on Set5, Urban100, and Manga109 datasets with the scale factor $\times2$. It is experimentally observed that the CA mechanism provides a generally effective technique for the fusion of features extracted in different directions and at various convolutional stages. For instance, the PSNR index increases from 38.03 dB to 38.11 dB on the Manga109 dataset when incorporating a LCA mechanism within each individual OAMs. The index is further boosted from 38.11 dB to 38.38 dB by utilizing the GCA mechanism to re-calculate channel-wise weights for the concatenated hierarchical features. The underlying principle is that LCA/GCA mechanisms can adaptively assign higher weights for the informative feature channels as well as to suppress redundant ones to generate more informative fused features and achieve higher SISR accuracy. 

\begin{figure}[ht]
		\begin{minipage}[t]{0.9\linewidth}
		\includegraphics[width=1\linewidth]{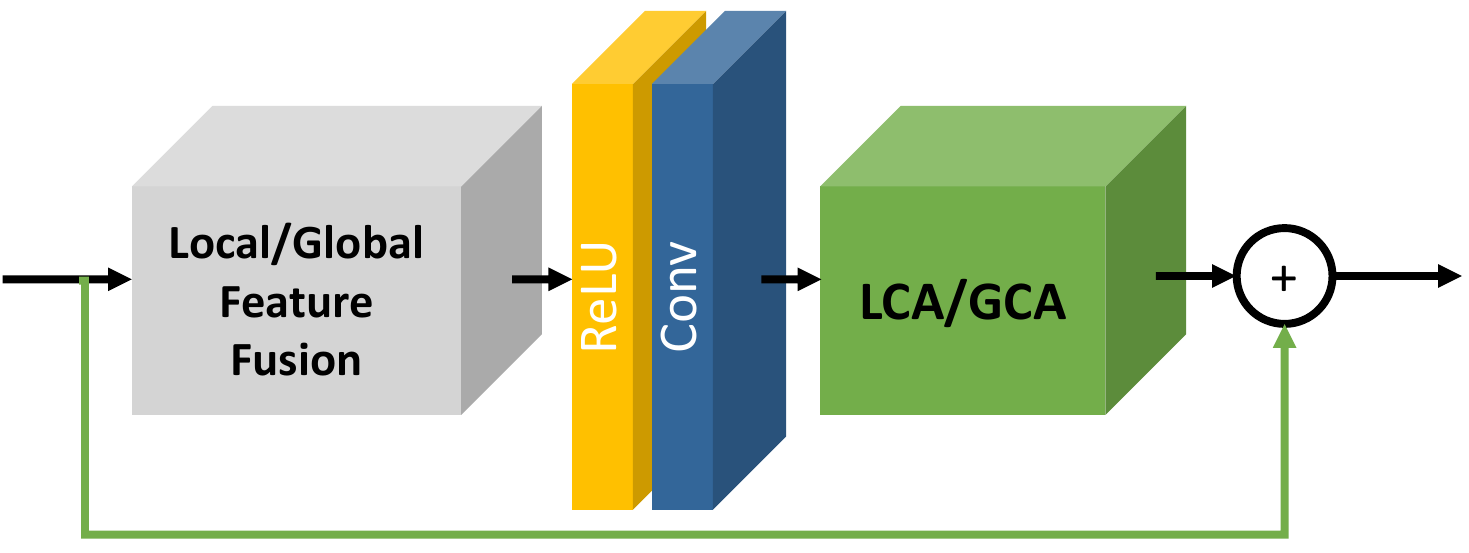}
		\centering{(a)}\\
	\end{minipage}
			\begin{minipage}[t]{0.9\linewidth}
		\includegraphics[width=1\linewidth]{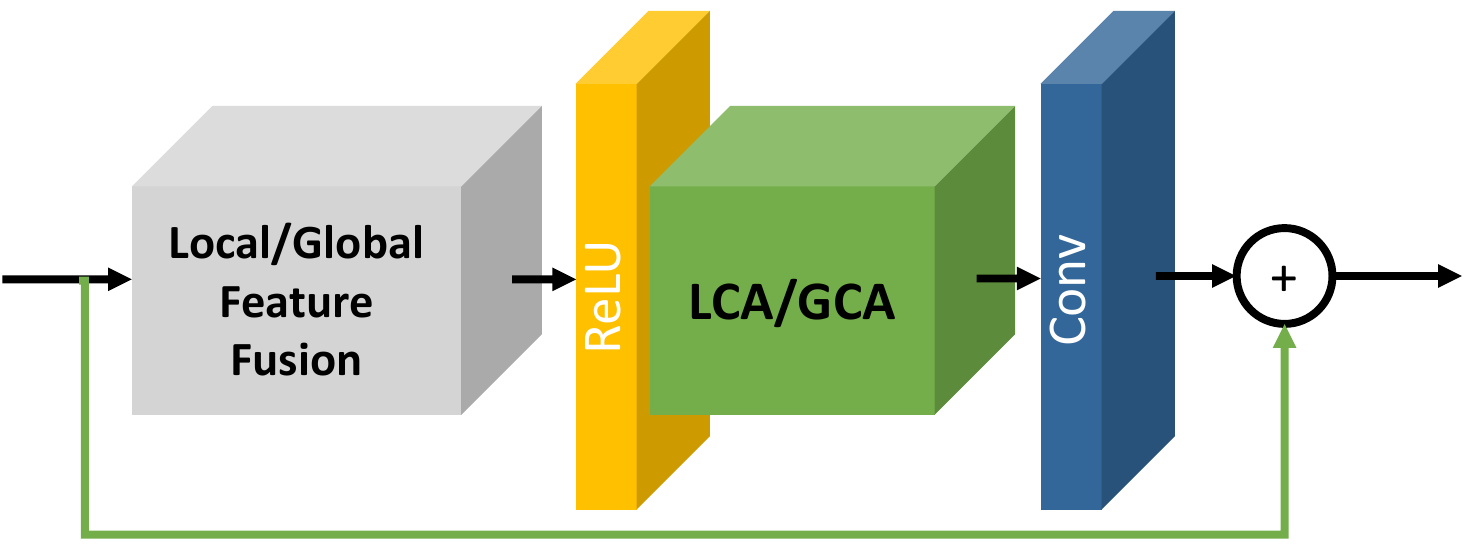}
		\centering{(b)}
	\end{minipage}\\
		\begin{minipage}[t]{0.9\linewidth}
		\includegraphics[width=1\linewidth]{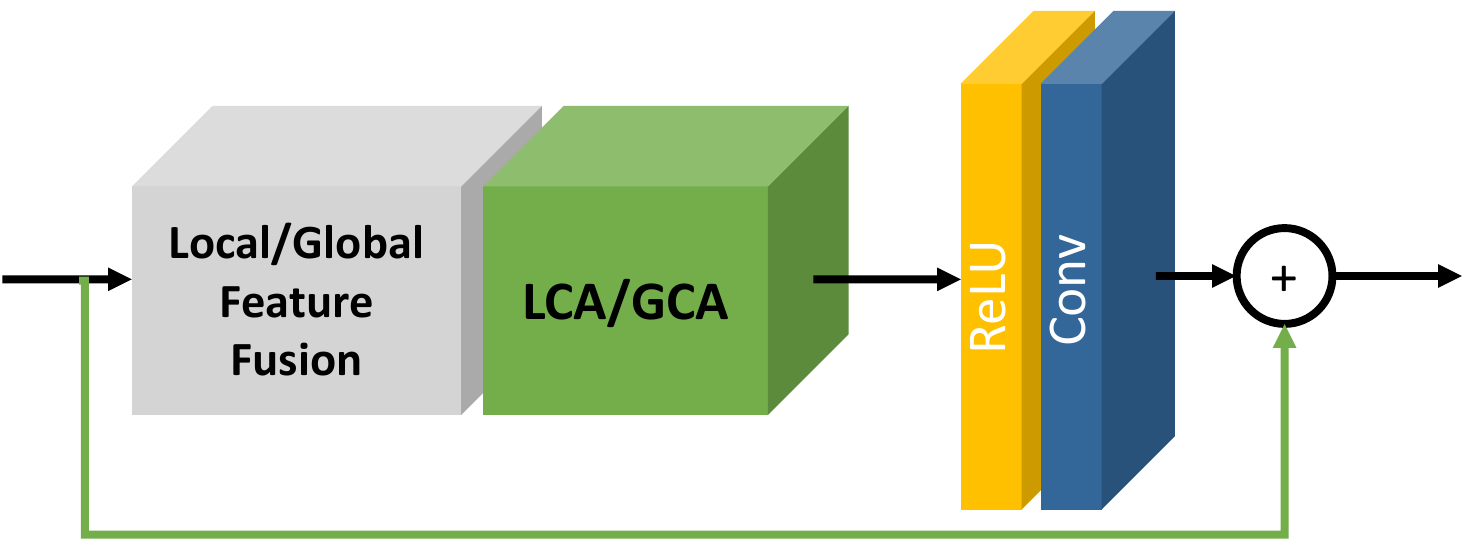}
		\centering{(c)}
	\end{minipage}
    \caption{Three design options in which the CA mechanisms are placed in different positions in a feature fusion module. (a) The CA-based re-calibration functions are applied after a ReLU activation function and a convolutional layer, (b) The CA-based re-calibration functions are placed between the ReLU and convolutional layers, (c) The LCA/GCA mechanisms are deployed before the ReLU and convolutional layers. Note Design (a) is the commonly adopted CA configuration in many SISR models \cite{CA2018BMVC,hu2019channel,zhang2018image,lu2018channel,cheng2018sesr}.}
	\label{fig:analysis2-2}
\end{figure}

Moreover, we experimentally evaluate a number of design options in which the CA mechanisms are placed in different positions in a feature fusion module. In Design (a) (Fig.~\ref{fig:analysis2-2} (a)), we place the LCA/GCA mechanisms after a ReLU activation function and a convolutional layer which is the commonly adopted configuration in many SISR models \cite{CA2018BMVC,hu2019channel,zhang2018image,lu2018channel,cheng2018sesr}. In Design (b) (Fig.~\ref{fig:analysis2-2} (b)), we put the CA re-calibration functions between the ReLU and convolutional layers. In Design (c) (Fig.~\ref{fig:analysis2-2} (c)), we move the LCA/GCA mechanism to the position before the ReLU and convolutional layers. Note the ReLU activation layer is utilized to embed more nonlinear terms into the network, and the convolutional layer compresses the channel number of concatenated features. Tab.~\ref{tab:tab10} shows the experimental results of different designs on Set5, Urban100, and Manga109 datasets with the scale factor $\times2$. It is observed that Design (c) achieves higher PSNR and SSIM indexes on all testing datasets than other alternatives. The experimental results illustrate that it is better to immediately utilize the CA mechanism to re-calibrate channel-wise weights for concatenated feature maps before squeezing the channel number of features (the convolutional layer) or converting the negative inputs to zeros (the ReLU activation function). 

\begin{table}[ht]
	\caption{Comparative evaluation of three design options in which the CA-based channel weights re-calibration is performed in different positions in a feature fusion module. The experimental metrics (PSNR(dB) and SSIM) are calculated on Set5, Urban100 and Manga109 datasets with scale factor $\times 2$.}
	\label{tab:tab10}
	\begin{center}
		\begin{tabular}{|cc||ccc|}
			\hline
			\multicolumn{2}{|c||}{\multirow{2}{*}{Comparison}} & \multicolumn{3}{c|}{Different CA-based Fusions} \\ 
			\multicolumn{2}{|c||}{\multirow{2}{*}{}}            & Design (a)     & Design (b)     & Design (c)    \\ \hline \hline
			\multirow{2}{*}{Set5}       & PSNR   & 37.91          & 37.93          & \textbf{37.97}         \\ 
			& SSIM      & 0.9603        & 0.9602         & \textbf{0.9605}        \\ \hline
			\multirow{2}{*}{Urban100}   & PSNR   & 31.46          & 31.50          & \textbf{31.57}        \\ 
			& SSIM      & 0.9220        & 0.9221         & \textbf{0.9226}       \\ \hline
			\multirow{2}{*}{Manga109}   & PSNR   & 38.13          & 38.21          & \textbf{38.38}        \\ 
			& SSIM      & 0.9751        & 0.9748         & \textbf{0.9755}       \\ \hline
		\end{tabular}
	\end{center}
\end{table}

\subsection{Comparisons with State-of-the-art SISR Methods}

Firstly, we compare our proposed light-weight SISR-CA-OA model (containing 10 OAMs) with a number of fast and accurate SISR methods which are also trained on the RGB91 \cite{yang2010image} and BSD \cite{martin2001database} datasets. More specifically, we consider Aplus \cite{timofte2014a+}, SelfExSR \cite{huang2015single}, SRCNN \cite{dong2016image}, VDSR \cite{kim2016accurate}, DRCN \cite{kim2016deeply}, ms-LapSRN \cite{lai2018fast}, DRRN \cite{tai2017image}, MemNet \cite{tai2017memnet}, and TSCN \cite{hui2018two}. Source codes or pre-trained models of these methods are publicly available.

\begin{figure*}[p]
	\small
	\centering
	\begin{minipage}{0.15\textwidth}
		\includegraphics[width=1\linewidth]{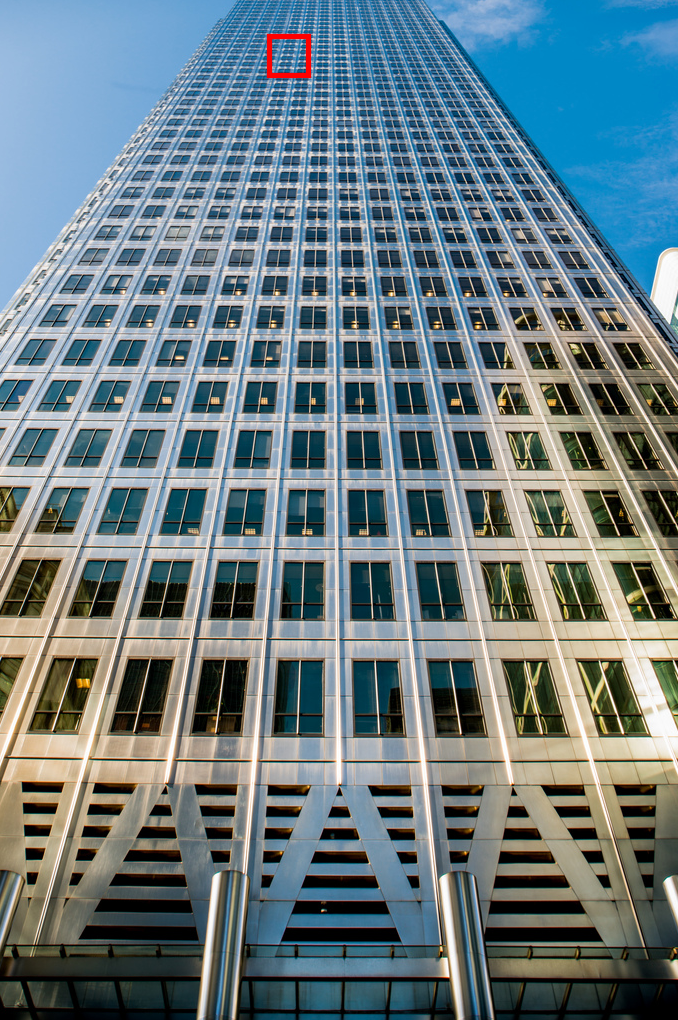}\\
		\centering{Urban100}\\
		\centering{img030($\times2$)}
	\end{minipage}
	\begin{minipage}{0.15\textwidth}
		\includegraphics[width=1\linewidth]{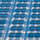}\\
		\centering{Ground Truth}\\
		\centering{PSNR/SSIM}\\
		\includegraphics[width=1\linewidth]{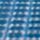}\\
		\centering{ms-LapSRN}\\
		\centering{30.28/0.9476}
	\end{minipage}
	\begin{minipage}{0.15\textwidth}
		\includegraphics[width=1\linewidth]{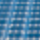}\\
		\centering{Bicubic}\\
		\centering{25.58/0.8569}\\
		\includegraphics[width=1\linewidth]{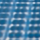}\\
		\centering{DRRN}\\
		\centering{30.07/0.9465}
	\end{minipage}
	\begin{minipage}{0.15\textwidth}
		\includegraphics[width=1\linewidth]{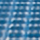}\\
		\centering{SRCNN}\\
		\centering{28.18/0.9148}\\
		\includegraphics[width=1\linewidth]{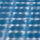}\\
		\centering{MemNet}\\
		\centering{30.15/0.9474}
	\end{minipage}
	\begin{minipage}{0.15\textwidth}
		\includegraphics[width=1\linewidth]{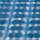}\\
		\centering{VDSR}\\
		\centering{29.71/0.8404}\\
		\includegraphics[width=1\linewidth]{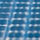}\\
		\centering{TSCN}\\
		\centering{\color{blue}30.33/0.9486}\\
	\end{minipage}
	\begin{minipage}{0.15\textwidth}
		\includegraphics[width=1\linewidth]{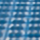}\\
		\centering{DRCN}\\
		\centering{29.87/0.9416}\\
		\includegraphics[width=1\linewidth]{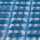}\\
		\centering{SISR-CA-OA}\\
		\centering{\color{red}30.38/0.9495}\\
	\end{minipage}
	\caption{Visual comparison of $\times 2$ SISR results for ``img030" in the Urban100 dataset. Note all SISR models are trained on the RGB91 \cite{yang2010image} and BSD \cite{martin2001database} datasets.}
	\label{fig:com1}
\end{figure*}

\begin{figure*}[p]
	\small
	\centering
	\begin{minipage}{0.15\textwidth}
		\includegraphics[width=1\linewidth]{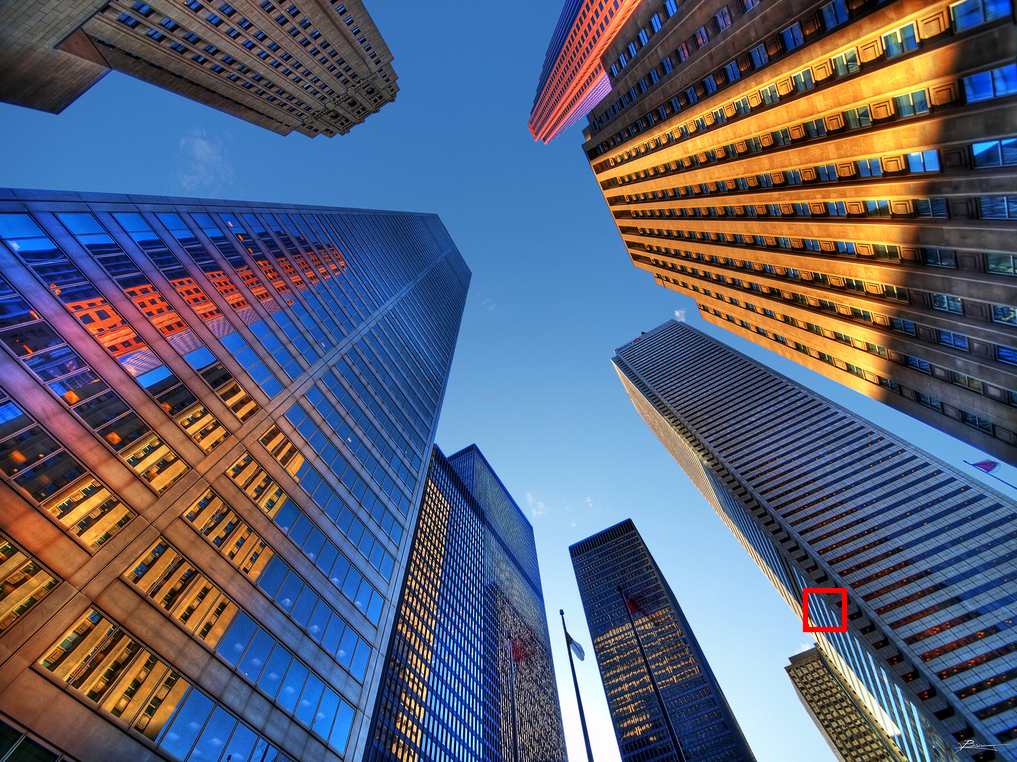}\\
		\centering{Urban100}\\
		\centering{img012($\times3$)}
	\end{minipage}
	\begin{minipage}{0.15\textwidth}
		\includegraphics[width=1\linewidth]{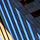}\\
		\centering{Ground Truth}\\
		\centering{PSNR/SSIM}\\
		\includegraphics[width=1\linewidth]{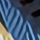}\\
		\centering{ms-LapSRN}\\
		\centering{24.52/0.7853}
	\end{minipage}
	\begin{minipage}{0.15\textwidth}
		\includegraphics[width=1\linewidth]{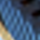}\\
		\centering{Bicubic}\\
		\centering{23.45/0.6894}\\
		\includegraphics[width=1\linewidth]{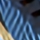}\\
		\centering{DRRN}\\
		\centering{24.69/0.7905}
	\end{minipage}
	\begin{minipage}{0.15\textwidth}
		\includegraphics[width=1\linewidth]{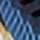}\\
		\centering{SRCNN}\\
		\centering{24.22/0.7521}\\
		\includegraphics[width=1\linewidth]{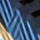}\\
		\centering{MemNet}\\
		\centering{{\color{blue}24.66/0.7910}}
	\end{minipage}
	\begin{minipage}{0.15\textwidth}
		\includegraphics[width=1\linewidth]{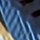}\\
		\centering{VDSR}\\
		\centering{24.47/0.7791/5.142}\\
		\includegraphics[width=1\linewidth]{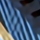}\\
		\centering{TSCN}\\
		\centering{24.63/0.7902}\\
	\end{minipage}
	\begin{minipage}{0.15\textwidth}
		\includegraphics[width=1\linewidth]{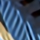}\\
		\centering{DRCN}\\
		\centering{24.47/0.7815}\\
		\includegraphics[width=1\linewidth]{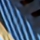}\\
		\centering{SISR-CA-OA}\\
		\centering{\color{red}24.72/0.7922}\\
	\end{minipage}
		\caption{Visual comparison of $\times 3$ SISR results for ``img012" in the Urban100 dataset. Note all SISR models are trained on the RGB91 \cite{yang2010image} and BSD \cite{martin2001database} datasets.}
	\label{fig:com2}
\end{figure*}

\begin{figure*}[p]
	\small
	\centering
	\begin{minipage}{0.15\textwidth}
		\includegraphics[width=1\linewidth]{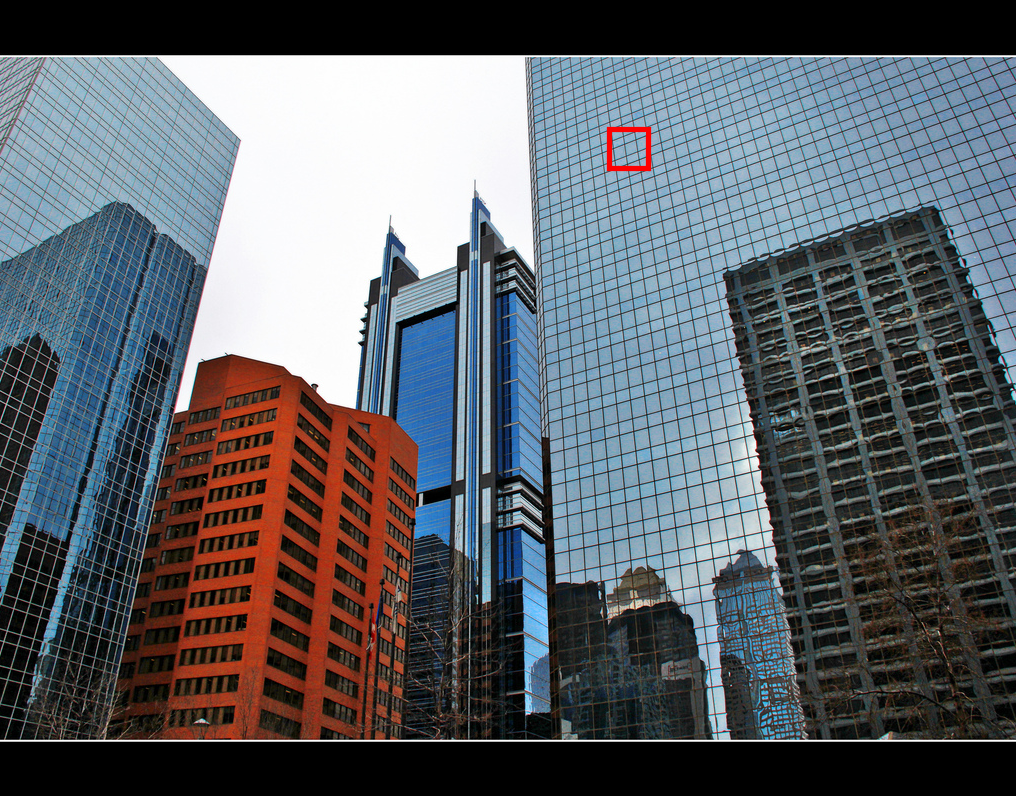}\\
		\centering{Urban100}\\
		\centering{img099($\times4$)}
	\end{minipage}
	\begin{minipage}{0.15\textwidth}
		\includegraphics[width=1\linewidth]{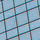}\\
		\centering{Ground Truth}\\
		\centering{PSNR/SSIM}\\
		\includegraphics[width=1\linewidth]{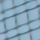}\\
		\centering{ms-LapSRN}\\
		\centering{\color{blue}25.36/0.7586}
	\end{minipage}
	\begin{minipage}{0.15\textwidth}
		\includegraphics[width=1\linewidth]{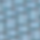}\\
		\centering{Bicubic}\\
		\centering{22.41/0.5873}\\
		\includegraphics[width=1\linewidth]{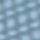}\\
		\centering{DRRN}\\
		\centering{25.09/0.7444}
	\end{minipage}
	\begin{minipage}{0.15\textwidth}
		\includegraphics[width=1\linewidth]{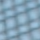}\\
		\centering{SRCNN}\\
		\centering{23.77/0.6714}\\
		\includegraphics[width=1\linewidth]{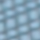}\\
		\centering{MemNet}\\
		\centering{25.16/0.7493}
	\end{minipage}
	\begin{minipage}{0.15\textwidth}
		\includegraphics[width=1\linewidth]{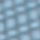}\\
		\centering{VDSR}\\
		\centering{24.02/0.6961}\\
		\includegraphics[width=1\linewidth]{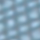}\\
		\centering{TSCN}\\
		\centering{25.08/0.7520}\\
	\end{minipage}
	\begin{minipage}{0.15\textwidth}
	    \includegraphics[width=1\linewidth]{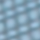}\\
		\centering{DRCN}\\
		\centering{23.71/0.6865}\\
		\includegraphics[width=1\linewidth]{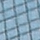}\\
		\centering{SISR-CA-OA}\\
		\centering{\color{red}25.39/0.7628}\\
	\end{minipage}
			\caption{Visual comparison of $\times 4$ SISR results for ``img099" in the Urban100 dataset. Note all SISR models are trained on the RGB91 \cite{yang2010image} and BSD \cite{martin2001database} datasets.}
	\label{fig:com3}
\end{figure*}

Tab.~\ref{tab:tab5} (PSNR and SSIM indexes) show quantitative evaluation results on Set5, Set14, B100, Urban100, and Manga109 with the scale factors $\times2$, $\times3$, $\times4$. Tab.~\ref{tab:tab7} shows the averaged running time of different SISR methods to process 100 input images of three different resolutions including $480\times360$, $640\times480$ and $1280\times720$. The testing is conducted on a PC which is equipped with NVIDIA Quadro P6000 GPU (24 GB memory). It is observed that our proposed SISR-CA-OA model performs favorably against these SISR models in terms of both restoration accuracy and computational efficiency. It achieves higher PSNR and SSIM values than some very deep networks (e.g., DRCN \cite{kim2016deeply}, DRRN \cite{tai2017image}, MemNet \cite{tai2017memnet}) and runs faster compared with some light-weight SISR models such as TSCN \cite{hui2018two} and ms-LapSRN \cite{lai2018fast}. Some visual comparative results with state-of-the-art deep-learning-based SISR methods are shown in Fig.~\ref{fig:com1}, \ref{fig:com2}, and \ref{fig:com3}. It is observed that our SISR-CA-OA model can achieve better image restoration results for three different scale factors ($\times2$, $\times3$, and $\times4$). As shown in Fig.~\ref{fig:com1} and Fig.~\ref{fig:com3}, the SISR-CA-OA model can restore sharper and clearer texture patterns in the highlighted regions. Moreover, it can effectively suppress undesired artifacts or distortions when reconstructing parallel edges/structures, as illustrated in Fig.~\ref{fig:com2}.

\begin{figure*}[ht]
	\centering
	\begin{minipage}[t]{0.18\linewidth}
		\includegraphics[width=1\linewidth]{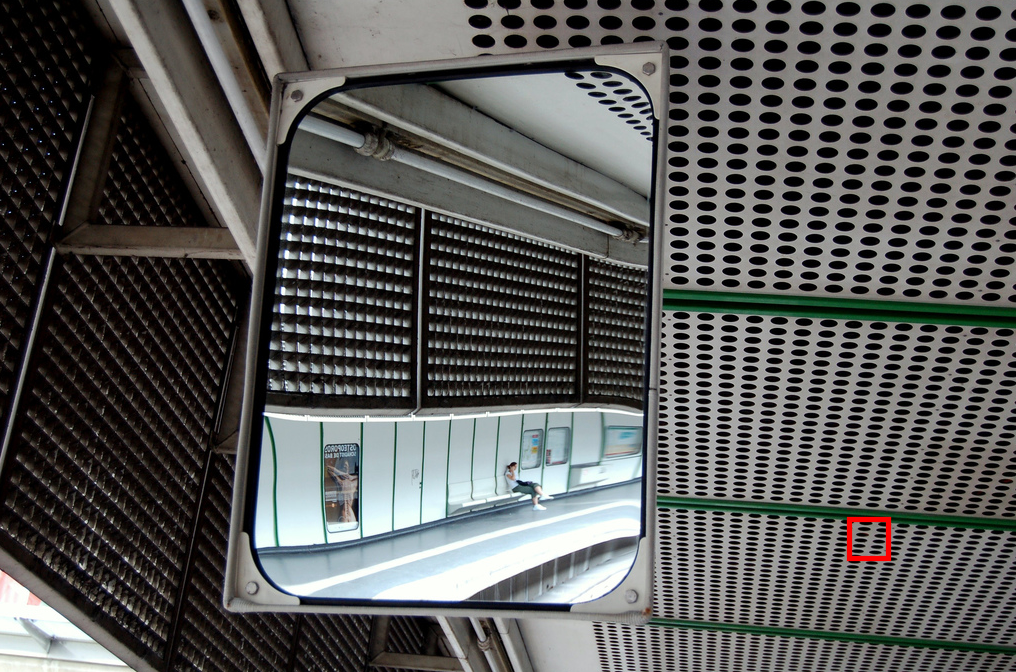}
		\centering{Urban100} \\
		\centering{img004($\times4$)}
	\end{minipage}
	\begin{minipage}[t]{0.12\linewidth}
		\includegraphics[width=1\linewidth]{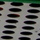}
		\centering{Ground Truth}\\
		\centering{PSNR/SSIM}\\
	\end{minipage}
	\begin{minipage}[t]{0.12\linewidth}
		\includegraphics[width=1\linewidth]{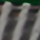}
		\centering{MSRN}\\
		\centering{23.80/0.8432}\\
	\end{minipage}
	\begin{minipage}[t]{0.12\linewidth}
		\includegraphics[width=1\linewidth]{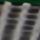}
		\centering{D-DBPN}\\
		\centering{23.74/0.8491}\\
	\end{minipage}
	\begin{minipage}[t]{0.12\linewidth}
		\includegraphics[width=1\linewidth]{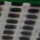}
		\centering{EDSR}\\
		\centering{{\color{blue}24.20}/0.8605}
	\end{minipage}
	\begin{minipage}[t]{0.12\linewidth}
		\includegraphics[width=1\linewidth]{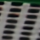}
		\centering{RDN}\\
		\centering{24.13/{\color{blue}0.8628}}
	\end{minipage}
	\begin{minipage}[t]{0.12\linewidth}
		\includegraphics[width=1\linewidth]{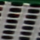}
		\centering{SISR-CA-OA$\ast$}\\
		\centering{{\color{red}25.34}/\color{red}0.8783}
	\end{minipage}
	\caption{Visual comparison of $\times 4$ SISR results for ``img004" in the Urban100 dataset. Note all SISR models are trained on the DIV2K dataset \cite{DIV2Kdataset}.}
	\label{fig:com5}
\end{figure*}

\begin{figure*}[ht]
	\centering
	\begin{minipage}[t]{0.18\linewidth}
		\includegraphics[width=1\linewidth]{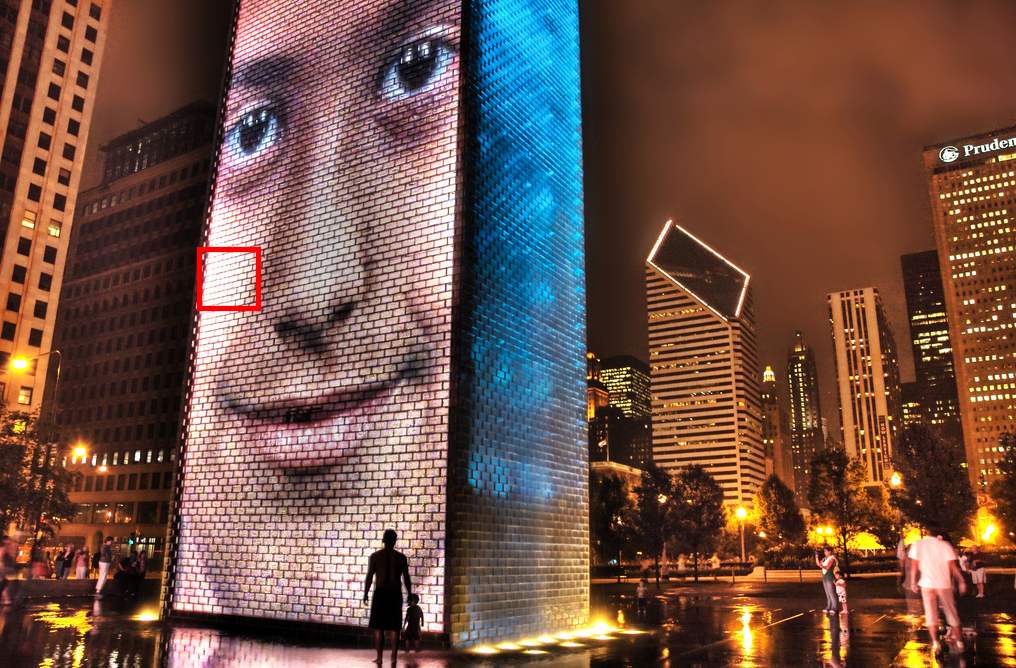}
		\centering{Urban100} \\
		\centering{img076($\times4$)}
	\end{minipage}
	\begin{minipage}[t]{0.12\linewidth}
		\includegraphics[width=1\linewidth]{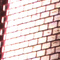}
		\centering{Ground Truth}\\
		\centering{PSNR/SSIM}\\
	\end{minipage}
	\begin{minipage}[t]{0.12\linewidth}
		\includegraphics[width=1\linewidth]{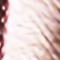}
		\centering{MSRN}\\
		\centering{23.06/0.7395}\\
	\end{minipage}
	\begin{minipage}[t]{0.12\linewidth}
		\includegraphics[width=1\linewidth]{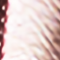}
		\centering{D-DBPN}\\
		\centering{23.18/ 0.7430}\\
	\end{minipage}
	\begin{minipage}[t]{0.12\linewidth}
		\includegraphics[width=1\linewidth]{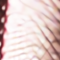}
		\centering{EDSR}\\
		\centering{23.90/0.7718}
	\end{minipage}
	\begin{minipage}[t]{0.12\linewidth}
		\includegraphics[width=1\linewidth]{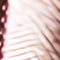}
		\centering{RDN}\\
		\centering{\color{blue}24.06/0.7791}
	\end{minipage}
	\begin{minipage}[t]{0.12\linewidth}
		\includegraphics[width=1\linewidth]{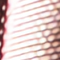}
		\centering{SISR-CA-OA$\ast$}\\
		\centering{\color{red}24.32/0.7877}
	\end{minipage}
	\caption{Visual comparison of $\times 4$ SISR results for ``img076" in the Urban100 dataset. Note all SISR models are trained on the DIV2K dataset \cite{DIV2Kdataset}.}
		\label{fig:com6}
\end{figure*}

Moreover, we compare the enhanced SISR-CA-OA$\ast$ model (containing 64 OAMs) with the best-performing SISR models trained on the high-resolution DIV2K dataset including MSRN \cite{li2018multi}, D-DBPN \cite{haris2018deep}, EDSR \cite{lim2017enhanced}, and RDN \cite{zhang2018residual}. The pre-trained models of these methods are publicly available. We calculate the average PSNR and SSIM values for scale factors $\times2$, $\times3$ and $\times4$ on Set5, Set14, B100, Urban100 and Manga109 testing datasets. As illustrated in Tab.~\ref{tab:tab8}, the proposed SISR-CA-OA$\ast$ model also achieves the highest PSNR and SSIM values in most cases. Compared with other SISR models trained the high-resolution DIV2K dataset, our SISR-CA-OA$\ast$ can more accurately restore complex image details (Fig.~\ref{fig:com5}) without incurring undesired artifacts (Fig.~\ref{fig:com6}) in large scale factor ($\times4$) SISR tasks.

\begin{table*}[ht]
	\caption{Benchmark results on state-of-the-art SISR methods. We calculate the average PSNR(dB)/SSIM values on Set5, Set14, B100, Urban100 and Manga109 datasets with scale factors $\times 2$, $\times 3$ and $\times 4$. The color {\color{red}red} and {\color{blue}blue} indicate the best and the second best performance respectively. It is noted that the metrics are calculated on Y channel (illumination channel of YCbCr color space).}
	\label{tab:tab5}
	\begin{center}
		\begin{tabular}{|c|c|ccccc|}
			\hline
			\multirow{2}{*}{Scale} & \multirow{2}{*}{Method} & Set5         & Set14        & B100       & Urban100      & Manga109    \\  
			&                         & PSNR / SSIM & PSNR / SSIM & PSNR / SSIM & PSNR / SSIM & PSNR / SSIM \\ \hline \hline
			\multirow{12}{*}{$\times 2$}   & Bicubic                    & 33.66 / 0.9299  & 30.24 / 0.8688  & 29.56 / 0.8431  & 26.88 / 0.8403  & 30.81 / 0.9341  \\ 
			& Aplus \cite{timofte2014a+}              & 36.54 / 0.9544  & 32.28 / 0.9056  & 31.21 / 0.8863  & 29.20 / 0.8938  & 35.37 / 0.9680  \\  
			& SelfExSR \cite{huang2015single}         & 36.50 / 0.9536  & 32.22 / 0.9034  & 31.17 / 0.8853  & 29.52 / 0.8965  & 35.12 / 0.9660  \\  
			& SRCNN \cite{dong2016image}              & 36.66 / 0.9542  & 32.45 / 0.9067  & 31.36 / 0.8879  & 29.51 / 0.8964  & 35.60 / 0.9663  \\  
			& VDSR \cite{kim2016accurate}             & 37.53 / 0.9587  & 33.03 / 0.9124  & 31.90 / 0.8960  & 30.76 / 0.9140  & 37.15 / 0.9738  \\  
			& DRCN \cite{kim2016deeply}               & 37.63 / 0.9588  & 33.04 / 0.9118  & 31.85 / 0.8942  & 30.75 / 0.9133  & 37.63 / 0.9740  \\ 
			& ms-LapSRN \cite{lai2018fast}               & 37.70 / 0.9590  & 33.25 / 0.9138  & 32.02 / 0.8970  & 31.13 / 0.9180  & 37.71 / 0.9747  \\ 
			& DRRN \cite{tai2017image}         & 37.74 / 0.9591  & 33.23 / 0.9136  & 32.05 / 0.8973  & 31.23 / 0.9188  & 37.88 / 0.9749  \\ 
			& MemNet \cite{tai2017memnet}             & 37.78 / 0.9597  & {\color{blue}33.28} / 0.9142  & 32.08 / 0.8978  & {\color{blue}31.31} / 0.9195  & 38.03 / {\color{red}0.9755}  \\  
			& TSCN \cite{hui2018two}                  & \color{blue}37.88 / 0.9602  & \color{blue}33.28 / 0.9147  & \color{blue}32.09 / 0.8985  & 31.29 / \color{blue}0.9198  & {\color{blue}38.07} / 0.9750  \\ 
			& SISR-CA-OA          & \color{red}37.97 / 0.9605  & \color{red}33.42 / 0.9158  & \color{red}32.15 / 0.8993  & \color{red}31.57 / 0.9226  & \color{red}38.38 / 0.9755  \\  \hline \hline
			\multirow{12}{*}{$\times 3$}   & Bicubic                 & 30.39 / 0.8682  & 27.55 / 0.7742  & 27.21 / 0.7385  & 24.46 / 0.7349  & 26.96 / 0.8546  \\  
			& Aplus \cite{timofte2014a+}              & 32.58 / 0.9088  & 29.13 / 0.8188  & 28.29 / 0.7835  & 26.03 / 0.7973  & 29.93 / 0.8120  \\  
			& SelfExSR \cite{huang2015single}         & 32.64 / 0.9097  & 29.15 / 0.8196  & 28.29 / 0.7840  & 26.46 / 0.8090  & 29.61 / 0.9050  \\ 
			& SRCNN \cite{dong2016image}              & 32.75 / 0.9090  & 29.29 / 0.8215  & 28.41 / 0.7863  & 26.24 / 0.7991  & 30.48 / 0.9117  \\  
			& VDSR \cite{kim2016accurate}             & 33.66 / 0.9213  & 29.77 / 0.8314  & 28.82 / 0.7976  & 27.14 / 0.8279  & 32.00 / 0.9329  \\ 
			& DRCN \cite{kim2016deeply}               & 33.82 / 0.9226  & 29.76 / 0.8311  & 28.80 / 0.7963  & 27.15 / 0.8276  & 32.31 / 0.9360  \\  
			& ms-LapSRN \cite{lai2018fast}               & 34.06 / 0.9249  & 29.97 / {\color{blue}0.8353}  & 28.92 / 0.8006  & 27.47 / 0.8369  & 32.68 / 0.9385  \\  
			& DRRN \cite{tai2017image}         & 34.03 / 0.9244  & 29.96 / 0.8349  & 28.95 / 0.8004  & 27.53 / {\color{blue}0.8378}  & 32.71 / 0.9379  \\   
			& MemNet \cite{tai2017memnet}             & 34.09 / 0.9248  & 30.00 / 0.8350  & {\color{blue}28.96} / 0.8001    & {\color{blue}27.56} / 0.8376        & \color{blue}32.79 / 0.9388  \\  
			& TSCN \cite{hui2018two}                  & \color{blue}34.18 / 0.9256  & {\color{blue}29.99} / 0.8351  & 28.95 / {\color{blue}0.8012}  & 27.46 / 0.8362  & 32.68 / 0.9381  \\  
			& SISR-CA-OA              & \color{red}34.23 / 0.9261  & \color{red}30.05 / 0.8363  & \color{red}29.01 / 0.8023  & \color{red}27.67 / 0.8403  & \color{red}32.92 / 0.9391  \\ \hline \hline
			\multirow{12}{*}{$\times 4$}   & Bicubic                 & 28.42 / 0.8104  & 26.00 / 0.7027  & 25.96 / 0.6675  & 23.14 / 0.6577  & 24.91 / 0.7846  \\ 
			& Aplus \cite{timofte2014a+}              & 30.28 / 0.8603  & 37.32 / 0.7491  & 26.82 / 0.7087  & 24.32 / 0.7183  & 27.03 / 0.8510  \\  
			& SelfExSR \cite{huang2015single}         & 30.30 / 0.8620  & 27.38 / 0.7516  & 26.84 / 0.7106  & 24.80 / 0.7377  & 26.80 / 0.8410  \\  
			& SRCNN \cite{dong2016image}              & 30.48 / 0.8628  & 27.50 / 0.7513  & 26.90 / 0.7103  & 24.52 / 0.7226  & 27.58 / 0.8555  \\  
			& VDSR \cite{kim2016accurate}             & 31.35 / 0.8838  & 28.02 / 0.7678  & 27.29 / 0.7252  & 25.18 / 0.7525  & 28.88 / 0.8854  \\  
			& DRCN \cite{kim2016deeply}               & 31.53 / 0.8854  & 28.03 / 0.7673  & 27.24 / 0.7233  & 25.14 / 0.7511  & 28.98 / 0.8870  \\  
			& ms-LapSRN \cite{lai2018fast}               & 31.72 / 0.8891  & 28.25 / 0.7730  & {\color{blue}27.42} / 0.7296  & \color{blue}25.50 / 0.7661  & 29.53 / {\color{blue}0.8956}  \\  
			& DRRN \cite{tai2017image}         & 31.68 / 0.8888  & 28.21 / 0.7720  & 27.38 / 0.7284  & 25.44 / 0.7638  & 29.44 / 0.8941  \\  
			& MemNet \cite{tai2017memnet}             & 31.74 / 0.8893  & 28.26 / 0.7723  & 27.40 / 0.7281  & {\color{blue}25.50} / 0.7630  & \color{red}29.64 / 0.8967  \\  
			& TSCN \cite{hui2018two}                  & {\color{blue}31.82} / {\color{red}0.8907}  & \color{blue}28.28 / 0.7734  & \color{blue}27.42 / 0.7301  & 25.44 / 0.7644  & 29.48 / 0.8954  \\  
			& SISR-CA-OA   & {\color{red}31.88} / {\color{blue}0.8900}  & \color{red}28.31 / 0.7740  & \color{red}27.45 / 0.7303  & \color{red}25.56 / 0.7670  & {\color{blue}29.61} / 0.8944  \\   \hline
		\end{tabular}
	\end{center}
\end{table*}

\begin{table*}[ht]
	\caption{Average running time for scale factor $\times 4$ on three different resolution settings, including 480 $\times$ 360, 640 $\times$ 480, 1280 $\times$ 720. The testing is conducted on a PC which is equipped with NVIDIA Quadro P6000 GPU (24 GB memory). The color {\color{red}red} and {\color{blue}blue} indicate the best and the second best performance respectively. }
	\label{tab:tab7}
	\begin{center}
		\begin{tabular}{|c|c|c|c|c|c|c|c|c|}
			\hline
			Dataset                  & Resolution & VDSR \cite{kim2016accurate}   & DRCN \cite{kim2016deeply}   & DRRN \cite{tai2017image}  & MemNet \cite{tai2017memnet} & ms-LapSRN \cite{lai2018fast}  & TSCN \cite{hui2018two}  & SISR-CA-OA \\ \hline \hline
			\multirow{3}{*}{Time(s)} & 480 $\times$ 360    & 0.0292 & 0.4895 & 5.9494 & 8.3749 & 0.0317 & \color{blue}0.0235 & \color{red}0.0151 \\  
			& 640 $\times$ 480    & 0.0512 & 0.8687 & 9.8489 & 15.7984 & 0.0453 & \color{blue}0.0379 & \color{red}0.0221 \\  
			& 1280 $\times$ 720   & 0.1553 & 2.6031 & 31.7105 & 47.8494 & 0.1368 & \color{blue}0.1047 & \color{red}0.0507 \\ \hline
		\end{tabular}
	\end{center}
\end{table*}



\begin{table*}[ht]
	\caption{Benchmark results of SISR models trained on the high-resolution DIV2K dataset. Average PSNR and SSIM values are calculated for scale factors $\times2$, $\times3$ and $\times4$ on Set5, Set14, B100, Urban100 and Manga109 datasets. The color {\color{red}red} and {\color{blue}blue} indicate the best and the second best performance respectively.}
	\label{tab:tab8}
	\begin{center}
		\begin{tabular}{|c||c||c|c|c|c|c|}
		\hline
		Dataset                          & Scale         &  MSRN \cite{li2018multi}            & D-DBPN \cite{haris2018deep}           & EDSR \cite{lim2017enhanced}               & RDN \cite{zhang2018residual}                 & SISR-CA-OA$\ast$ \\ 
		\hline \hline
		\multirow{3}{*}{Set5}     & $\times 2$ & 38.08 / 0.9605 &38.09 / 0.9600  &38.11 / 0.9601  &\color{red}38.24 / 0.9614   & \color{blue}38.22 / 0.9613   \\  
		                                     & $\times 3$ & 34.38 / 0.9262 &-- \ --    &34.65 / 0.9282 &\color{blue}34.71 / 0.9296   & \color{red}34.73 / 0.9297 \\  
		                                     & $\times 4$ & 32.07 / 0.8903 &{\color{blue}32.47} / 0.8980   &32.46 / 0.8968 &\color{blue}32.47 / 0.8990   & \color{red}32.55 / 0.8994\\			
		 \hline\hline
		 \multirow{3}{*}{Set14}    & $\times 2$& 33.74 / 0.9170&33.85 / 0.9190      &{\color{blue}33.92} / 0.9195 &\color{red}34.01 / 0.9212    & 33.91 / \color{blue}0.9208 \\  
			                                   & $\times 3$& 30.34 / 0.8395&-- \ --      &30.52 / 0.8462 &\color{blue}30.57 / 0.8468    &  \color{red}30.59 / 0.8470\\ 
			                                   & $\times 4$& 28.60 / 0.7751& {\color{blue}28.82} / 0.7860    &28.80 / \color{blue}0.7876  &28.81 / 0.7871    &  \color{red}28.86 / 0.7882\\ 
		\hline \hline
		\multirow{3}{*}{B100}   & $\times 2$ &  32.23 / 0.9013& 32.27 / 0.9000&32.32 / 0.9013 &{\color{blue}32.34} / \color{red}0.9017     & {\color{red}32.35} / \color{blue}0.9016 \\ 
			                                & $\times 3$ & 29.08 / 0.8041&-- \ --   &29.25 / \color{blue}0.8093 & \color{blue}29.26 / 0.8093   & \color{red}29.29 / 0.8099\\  
			                                & $\times 4$ & 27.52 / 0.7273& 27.72 / 0.7400&27.71 / \color{blue}0.7420  & {\color{blue}27.72} / 0.7419     &\color{red}27.76 / 0.7424  \\ 
		\hline\hline
		\multirow{3}{*}{Urban100} & $\times 2$ &32.22 / 0.9326&32.55 / 0.9324&{\color{blue}32.93} / 0.9351 &32.89 / \color{blue}0.9353     & \color{red}33.03 / 0.9359 \\  
			                                     & $\times 3$ &28.08 / 0.8554&-- \ --   &\color{blue}28.80 / 0.8653 &\color{blue}28.80 / 0.8653     & \color{red}28.98 / 0.8680 \\  
			                                     & $\times 4$ &26.04 / 0.7896&26.38 / 0.7946&\color{blue}26.64 / 0.8033 & 26.61 / 0.8028     &  \color{red}26.74 / 0.8060\\ 
		\hline\hline
		\multirow{3}{*}{Manga109} & $\times 2$ & 38.82 / 0.9868&38.89 / 0.9775& 39.10 / 0.9773 & {\color{blue}39.18} / \color{red}0.9780     & {\color{red}39.24} / \color{blue}0.9778 \\  
			                                      & $\times 3$ &33.44 / 0.9427&-- \ --  & {\color{blue}34.17} / 0.9476 & 34.13 / \color{blue}0.9484     & \color{red}34.38 / 0.9493 \\ 
			                                      & $\times 4$ &30.17 / 0.9034&30.91 / 0.9137& {\color{blue}31.02} / 0.9148 & 31.00 / \color{blue}0.9151      & \color{red}31.22 / 0.9168 \\ 
		\hline
		\end{tabular}
	\end{center}
\end{table*}

\section{Conclusion}
\label{conclusion}

In this paper, we proposed a novel CNN-based model for high-quality SISR via channel attention-based fusion of orientation-aware features. Instead of utilizing square-shaped convolutional kernels (e.g., $3\times3$ or $5\times5$) to extract features \cite{dong2014learning, kim2016accurate, tai2017image, tong2017image, tai2017memnet}, we integrate multiple convolutional kernels of various shapes (i.e., $5\times1$, $1\times5$, and $3\times3$) in a single feature extraction module to extract orientation-aware features. Moreover, we adopt the channel attention mechanism for the local fusion of features extracted in different directions and the global fusion of features extracted in hierarchical stages. Extensive benchmark evaluations well demonstrate that our proposed SISR-CA-OA model achieves superiority over state-of-the-art SISR methods \cite{dong2016image, kim2016accurate, kim2016deeply, tai2017image, tai2017memnet, hui2018two, li2018multi, haris2018deep, lim2017enhanced, zhang2018residual} in terms of both restoration accuracy and computational efficiency. 

\appendices

\ifCLASSOPTIONcaptionsoff
  \newpage
\fi

\bibliographystyle{IEEEtran}

\bibliography{IEEEabrv,mybibfile}

\end{document}